# Theoretical and computational tools to model multistable gene regulatory networks

Federico Bocci[1,2,*], Dongya Jia[3,#], Qing Nie[1,2], Mohit Kumar Jolly[7], José Onuchic[3-6,*]

[1]The NSF-Simons Center for Multiscale Cell Fate Research; [2]Department of Mathematics, University of California, Irvine, CA 92697, USA

[3]Center for Theoretical Biological Physics, [4]Departments of Physics and Astronomy, [5]Chemistry and [6]Biosciences, Rice University, Houston, TX 77005, USA

[7]Centre for BioSystems Science and Engineering, Indian Institute of Science, Bangalore 560012, India

[#]Current address: Immunodynamics Group, Laboratory of Integrative Cancer Immunology, Center for Cancer Research, National Cancer Institute, Bethesda, MD 20814,USA

Corresponding authors:
*Federico Bocci: fbocci@uci.edu
*José Onuchic: jonuchic@rice.edu

## Abstract

The last decade has witnessed a surge of theoretical and computational models to describe the dynamics of complex gene regulatory networks, and how these interactions can give rise to multistable and heterogeneous cell populations. As the use of theoretical modeling to describe genetic and biochemical circuits becomes more widespread, theoreticians with mathematical and physical backgrounds routinely apply concepts from statistical physics, non-linear dynamics, and network theory to biological systems. This review aims at providing a clear overview of the most important methodologies applied in the field while highlighting current and future challenges. It also includes hands-on tutorials to solve and simulate some of the archetypical biological system models used in the field. Furthermore, we provide concrete examples from the existing literature for theoreticians that wish to explore this fast-developing field. Whenever possible, we highlight the similarities and differences between biochemical and regulatory networks and "classical" systems typically studied in non-equilibrium statistical and quantum mechanics.

**Table of Contents**

# 1. Introduction

Despite sharing similar genetic information, cells in living organisms can diversify into many different cell types, each of whom carrying a distinct and unique function [1]. For example, cells in the human body differentiate into more than 200 different cell types that can be distinguished based on the different profiles of molecular signatures [2]. The process of molecule production can be summarized at a very simplified level by stating that special sequences of the genome (genes) are used as templates to synthetize messenger RNA (mRNA) molecules, which are in turn translated into proteins by the cell's machinery. Yet, transcription (gene to mRNA) and translation (mRNA to protein) are not independent processes, but rather happen in a complex cellular environment. For example, special molecules called transcription factors can regulate the production of other molecular species by binding to the DNA in proximity to genes and either facilitating or obstructing transcription [3]. This seemingly simple mechanism can give rise to a spectacular level of complexity when considering that living cells have tens of thousands of genes, and their mutual interactions are only partially understood [4,5].

The purpose of this review is to provide an overview of the mathematical strategies used to describe this complex phenomenon and predict how cellular heterogeneity emerges from the underlying regulatory dynamics. The fundamental assumptions connecting all these modeling strategies is that it is possible to describe, with some extent of success, the intricated interactions between genes and transcription factors with mathematical models (such as differential equations or logic circuits), and that the outputs of these models (such as probability distributions or attractors) can be related to the different states that cells can attain. We acknowledge the existence of other recent reviews that discuss the application of mathematical and physical concepts to chemical-reaction systems, stochastic events, and simulation strategies in biological systems [6–9]. Compared to these reviews, which explore the theoretical aspects in deeper details, we aim at presenting a more hands-on and simplified overview of the main theoretical strategies necessary to investigate the multistability and heterogeneity in biological systems, explicitly discussing examples and offering hands-on tutorials and simulations. With this approach, we aim to create a unifying ground not only for theoreticians interested in biological systems, but also for biologists who are interested in the theoretical tools to model complex biological systems. In this sense, this review can be seen as a set of suggestions to tackle gene regulation and biochemical interactions at different levels of complexity.

In the first section of the review, we introduce theoretical and computational methods to model the emerging dynamics of gene regulatory networks and their contribution in many biological open problems. First, we discuss discrete, stochastic frameworks to model transcription, translation, and interactions between transcription factors. Further, we introduce a continuous framework that allows to model larger circuits with several transcription factors, and discuss how to characterize multistable systems with the pseudopotential landscape, the "systems biology" analogy to the potential landscape. Finally, we discuss strategies to tackle large circuits, where missing information about the model's parameters require approximation methods such as Boolean networks and parameter randomization, and provide an introduction to how gene regulatory networks can be inferred using high-resolution single cell sequencing data. This section also includes [hands-on tutorials](#) to simulate some of the simplest and most widely used circuit structures in the literature. In the second section of the review, we discuss in detail three specific biological examples where the application of these methods led to new and significant biological insights, including the epithelial-mesenchymal transition, differentiation of stem cells, and cell-cell communication through Notch signaling. These examples, which were chosen not least due to the expertise of the authors, cannot fully do justice to the extensive available literature, and we would like to acknowledge the many outstanding works that could not be included here due to the limitation of space. While discussing these three examples, we introduce further mathematical tools that have been especially important in their respective fields, including: the modeling of non-coding RNA in the epithelial-mesenchymal transition, the application of quantum many-body formalism in the modeling of stem cell differentiation, and spatial models with many cells in the description of cell-cell communication through Notch signaling.

## 2. Mathematical methods to study the multistability of regulatory networks

**2.1 Discrete models of circuits: from reactions to the Chemical Master Equation**

2.1.1 <u>Construction of the CME for protein synthesis</u>

In this first section, we start by analyzing theoretical approaches that aim at modeling the stochastic nature of individual molecular reactions. The basic steps in molecule production include transcription, where a messenger RNA (mRNA) transcript is created, and translation, when the final molecule is produced from the mRNA transcript. This working model can be formalized by assuming that a gene transcribes mRNA molecules with constant rate $k_m$. Moreover, each mRNA molecule translates into proteins with rate constant $k_p$. Further, if mRNA and protein molecules degrade with rate constants $\gamma_m, \gamma_p$, we can write the following set of reactions

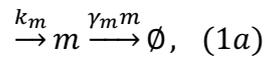
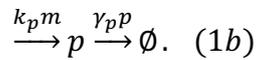

$$\xrightarrow{k_m} m \xrightarrow{\gamma_m m} \emptyset, \quad (1a)$$
$$\xrightarrow{k_p m} p \xrightarrow{\gamma_p p} \emptyset. \quad (1b)$$

Reactions (1a-b) are summarized in figure 1A. Starting from copy numbers of mRNA and protein $(m, p)$ at time $t$, the next reaction will move the system to a new configuration $(m', p')$ as shown in figure 1B. Since the rates in the reactions (1a-b) are either constant or depend linearly on $(m, p)$, the copy number of mRNA or protein can be increased or decreased only by one molecule at a time. This is not necessarily the case; in more complex processes molecules might bind or degrade as oligomer complexes, which cannot be described by linear rates; we will discuss this case in more details at a later point when introducing Hill functions. It is also interesting to note that, in eqs. (1a-b), four rates control the copy number of two molecular species. Therefore, infinite combinations of the parameters $(k_m, k_p, \gamma_m, \gamma_p)$ can give rise to the same $(m, p)$ state. Recent studies using high throughput data suggest, however, that only certain parameter combinations are naturally observed as a result of balancing between transcription precision, which is achieved with large rate constants, and costs, which are minimized by low rate constants [10].

Starting from reactions (1a-b), we aim at constructing an equation to predict the probability to observe a certain combination of copy numbers $(m, p)$ at any given time. To construct the Chemical Master equation (CME), we define a discrete, infinite set of probabilities $\{P(m, p, t)\}$, so that $P(m, p, t)$ describes the probability to observe $m$ mRNA molecules and $p$ protein molecules at time $t$. These probabilities satisfy a normalization condition (i.e., they sum up to one at any time $t$)

$$\sum_{n=0}^{+\infty} \sum_{p=0}^{+\infty} P(m, p, t) = 1. \quad (2)$$

Following the reaction scheme of Figure 1B, we can write an evolution equation for the set of probabilities

$$\frac{dP(m,p)}{dt} = k_m[P(m-1,p) - P(m,p)] + \gamma_m[(m+1)P(m+1,p) - mP(m,p)]$$
$$+ k_p m[P(m,p-1) - P(m,p)] + \gamma_p[(p+1)P(m,p+1) - pP(m,p)]. \quad (3)$$

In eq. (3), all the positive and negative terms represent influx and outflux of probability. In other words, positive terms in eq. (3) are associated with reactions that bring the system to the $(m, p)$ state, whereas negative terms are associated with reactions that bring the system out of the $(m, p)$ state (see again the reactions in Fig. 1B). Moreover, eq. (3) can be further simplified with reasonable assumptions when different terms on the right-hand size are associated to well-separated timescales. Specifically, one common approximation is to assume a quasi steady state approximation (QSSA) for mRNA, whereby the variable $m$ is substituted by its average value $k_m/\gamma_m$. In other words, the number of mRNAs can be treated as a constant if it equilibrates much faster than the protein. This assumption trades a more faithful description of the biological system for a reduction in complexity and computational time, and may or may not be accurate depending on the specific biological systems. For example, transcription and translation have similar rates in bacteria (about 1 min/gene and 1 min/protein), thus making QSSA a poor approximation. Translation, however, is significantly slower in mammalian cells (about 10 min/protein) due to an intermediate reaction called "RNA splicing", which will be discussed later in section 2.1.5, thus justifying transcription QSSA [11]. For comparison, protein loss due to degradation and/or cell division (typically referred to as "dilution") has a typical timescale of hours to days depending on the organism[11], thus making it the slower reaction in eq. (3). The simplified CME after mRNA QSSA describes only the protein copy number $p$

$$\frac{dP(p)}{dt} = +k[P(p-1) - P(p)] + \gamma_p[(p+1)P(p+1) - pP(p)], \quad (4)$$

where $k = k_p m = k_p k_m/\gamma_m$. Eq. (4) describes a birth-death process with rate constants ($k$, $\gamma_p$). The steady state solution of eq. (4) can be evaluated with different analytical, numerical or simulation methods that are explored in detail in the following section.

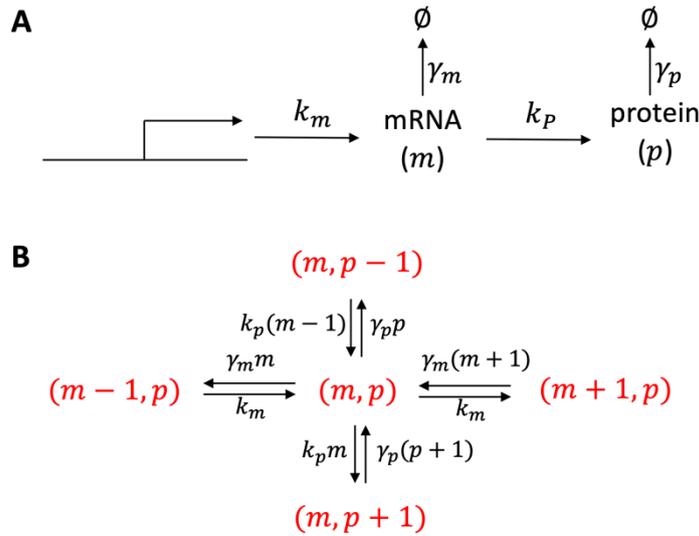

*Figure 1. **Circuit and reactions for transcription-translation chain.** (A) mRNA molecules are transcribed from the promoter with constant rate $k_m$. Each mRNA molecule is translated into proteins with rate constant $k_p$. Both mRNA and protein degrade with degradation rates $\gamma_m$, $\gamma_p$. (B) Schematic representation of all the possible reactions leading to or out of the configuration with $m$ mRNA transcripts and $p$ proteins.*

### 2.1.2 Solving and simulating the birth-death CME

The steady state solution of the CME (eq. 4) is the complete set $\{P_n^{SS}\}$ obtained by solving $\frac{dP_n^{SS}}{dt} = 0$. In this section, we discuss three methods that may be suitable to tackle different types of CMEs using the birth-death process (eq. 4) as an example. The insight box 1 in section 2.1.4 offers the explicitly solution/implemented of the three methods. First, it is possible to find analytical solutions of the CME if when the number of variables is small and the reactions are zeroth or first order. Typically, one first takes advantage of the fact that the equation for $P_0^{SS}$ has a simpler form since $n$ is strictly semipositive. Then, all the other $P_n^{SS}$ can be found iteratively. This method can be successfully applied to the birth-death process, yielding $P_n = \frac{1}{n!}\left(\frac{k}{\gamma}\right)^n e^{-\frac{k}{\gamma}}$ (see insight box 1).

Second, generating functions provide a more general framework to solve CMEs by converting the infinite-dimensional system of equations into the single differential equation. In the simple case of the birth-death process, this approach can be developed analytically (see insight box 1). Nonetheless, large number of variables and higher-order interaction terms typically prevent analytical solutions and potentially make numerical approaches expansive.

Finally, a third possibility to study the CME is to explicitly simulate the dynamics using algorithms such as non-rejective Monte Carlo, more commonly known as Gillespie algorithm [12,13]. In the Gillespie simulation scheme, chemical reactions are treated as independent events (see for example [14,15]). Under this assumption, the waiting time until the next realization of any given reaction with rate $w$ is an exponentially distributed random variable with mean and standard deviation equal to $1/w$. Therefore, an iteration of a Gillespie simulation with $i = 1, \ldots, M$ reactions includes the following steps. First, compute the individual reaction rates $w_i$. Second, draw an exponentially distributed random variable ($\tau$) with rate parameter $W = \sum_{i=1}^{M} w_i$ that represents the waiting time until the next reaction occurs. Finally, draw a uniformly distributed random variable ($r$) to select which reaction occurs. The probability of each reaction is given by its relative weight $w_i/W$ (a pseudocode is presented in the insight box 1 for the birth-death process). The result of the simulation is a trajectory of the number of molecules $n(t)$ as a function of time starting from an initial condition specified by the user. In the long-time limit, the statistics of $n(t)$ will resemble closely and closely the steady-state distribution $P_n$ obtained with iterative and generating function methods (provided that these approaches are feasible for the specific system under study). The long-time limit is identified as a timescale that is substantially larger than any typical timescale in the reaction scheme, and therefore depends on the parameters of the model (see the insight box 1 for this specific calculation in the birth-death process).

The steady state solution of the simplified birth-death process modeled by eq. (4) is a Poisson distribution with mean ($\mu_p$) and variance ($\sigma_p$) both equal to $k_P/\gamma_P$. However, as discussed in more detail by Tsimring [7], a full model that takes into account mRNA dynamics leads to a broader protein copy number distribution with variance equal to

$$\sigma_p = \mu_p \left( \frac{k_p}{\gamma_p + \gamma_m} + 1 \right). \quad (5)$$

Experimentally, it is well known that mRNA and protein count distributions in cells can exhibit variances much larger than the predicted Poisson distribution of the birth-death CME [16]. This effect arises due to the so-called transcriptional bursting. Namely, promoters do not transcribe mRNA molecules at a constant rate, but rather switch between 'ON' and 'OFF' periods, thus leading to bursts of mRNA molecule production on a typical timescale of a few minutes [16], thus leading to broader copy number distributions. The contribution of mRNA dynamics on protein copy number dispersion can be quantified with Gillespie simulations. The protein copy number distribution over the course of a long simulation is narrower in the simpler mRNA QSSA model, compared to the full mRNA-protein model (Fig. 2A). The relation between width of the distribution and mRNA QSSA can be quantified by testing the performance at different mRNA dynamics speed. The average, expected value of mRNA copy number is determined by the model's parameters: $m^{ss} = k_m/\gamma_m$. Therefore, increasing both the mRNA production rate $k_m$ and degradation rate $\gamma_m$ by the same factor does not modify the expected mRNA copy number, but makes the mRNA equilibration dynamics faster, thus making mRNA QSSA more and more precise. As the value of these parameters is increased, the standard deviation of protein ($p$) copy number in Gillespie simulations decreases, and finally becomes comparable to the QSSA standard deviation when the mRNA dynamics is sufficiently fast (Fig. 2B). Further discussion of transcriptional bursting and mathematical attempts to quantitatively capture model it are reviewed in [7,16].

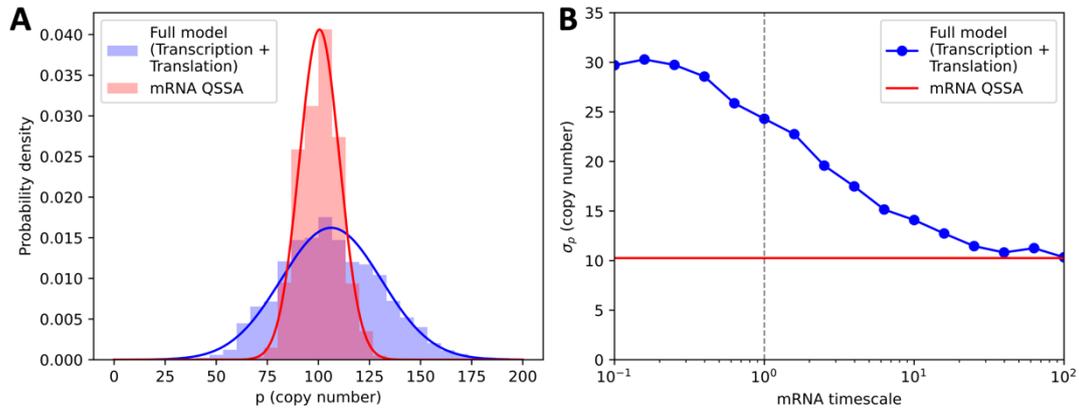

*Figure 2. **Testing the mRNA Quasi Steady State Approximation in protein synthesis.** (A) The distribution of protein copy number for the full model presented in eq. (3) (blue) and for mRNA QSSA corresponding to eq. (4) (red). Solid lines indicate Gaussian fit. (B) Blue: The standard deviation of protein (p) distribution for increasingly fast mRNA dynamics. The dashed black line indicates the parameters used in panel (A). Red: standard deviation for mRNA QSSA. The code to simulate protein synthesis is available in our protein synthesis tutorial.*

Both the full and QSSA models of protein synthesis result in a distribution with a single peak, which can be identified as the state assumed by the system, corresponding to a specific cell phenotype (see Fig. 2A). More complex systems can instead exhibit multimodal distributions, where the different peaks can be interpreted as the different, co-existing states/phenotypes available to a cell. By adding a self-activation loop, the probability distribution obtained with the Gillespie simulations becomes bimodal, with one peak corresponding to a low-expressing state and one peak corresponding to a high-expressing state (Fig. 3A). In the self-activation loop, the protein p acts as a transcription factor (TF) and activates the transcription of its own gene [17]. Furthermore, the simulated trajectory of protein copy number highlights transitions between the low-expressing and high-expressing states driven by the stochastic fluctuations (Fig. 3B). A detailed analysis of this circuit motif can be found in [17]. Despite its simplicity, the self-activating loop can describe realistic biological systems, as in the case of the cell fate differentiation in the developing fruit fly embryo based on the self-activation of the *ftz* gene, where the authors further test many parameter combinations and identify the conditions enabling bistability [18].

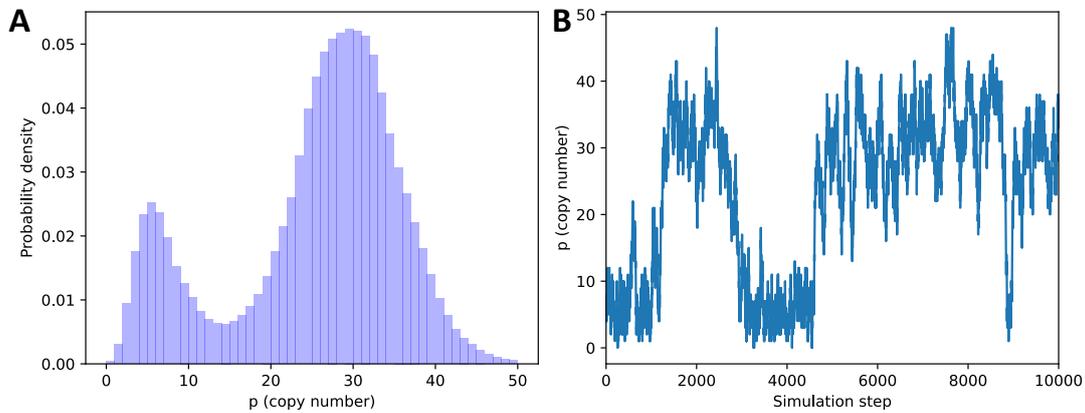

*Figure 3.* **Bimodal dynamics of the self-activating gene. (A)** *The bimodal copy number distribution in the self-activating gene.* **(B)** *A sample of simulation trajectory highlights the transitions between low expression and high expression.*

2.1.3 CME for the toggle switch and larger networks

The CME formalism can be extended to study the coupled dynamics of a network of interacting species. In this section, we will set up the CMEs to describe the toggle switch, a simple motif composed by two genes that mutually repress each other, which can be seen as a basic mechanism to model differentiation between two distinct cell types [19]. This method can serve as a general example to tackle more complex networks of interconnected biochemical species.

In the toggle switch, two genes (x and y) transcribe mRNA for the transcription factors (TF for brevity in the following) X and Y. We will follow the same strategy of the birth-death process of eq. (4) and assume QSSA for the mRNA species x and y to reduce the number of degrees of freedom and parameters. After being produced, TF X can inhibit the production of Y by binding to the promoter transcribing for gene y, and *vice versa* (see Figure 4A).

TFs do not necessarily bind and regulates the transcription of other species as single molecules; often, they first dimerize (or create higher-order oligomer complexes) and then bind to DNA. Here, we assume that both TFs X and Y dimerize with forward and backward rate constants $k_D, k_M$, as often assumed in existing mathematical modeling [20]. Molecular binding and unbinding that regulate TF dimerization have much shorter timescales compared to protein synthesis, thus providing a possibility to further simplify the model by assume QSSA for the dimerization/monomerization reactions[21] (see Figure 4A). Thus, the steady state concentrations of the dimers are $[X_2] = (k_D/k_M)X^2$ and $[Y_2] = (k_D/k_M)Y^2$. Following dimerization, the dimers bind and unbind to the DNA with rate constants $k_+, k_-$. Both TFs are produced at a rate $k_0$ when their respective promoters are not bound; the rate, however, decreases to $k_1 < k_0$ when the promoters are bound to the dimerized form of the TF. Therefore, the toggle switch can be viewed as two separate birth-death processes where the final molecular products X and Y bind to the other protein's gene to inhibit its production. All the considered reactions are depicted in Figure 4A.

The toggle switch can be characterized with four variables: two positive integer variables (n, m) accounting for the number of molecules for species X and Y, respectively, and two Boolean variables ($S_X, S_Y$) that encode the state of the two promoters. $S_i = 0$ implies that promoter of species $i$ is unbound, while $S_i = 1$ implies that promoter is bound, respectively. Thus, starting from a generic configuration $(n, m, S_x, S_y)$ at time $t$, a subsequent reaction can either modify the copy number of X and Y or change the state of one of the two promoters. All reactions and rates are illustrated in Figure 4B. Following the CME approach, we write an evolution equation for the probability $P(n, m, S_x, S_y)$. To write the CME in a compact form, we define the following functions. First, a generalized production rate function

$$k(S) = k_0(1 - S) + k_1(S). \quad (6)$$

This function assumes the value $k_0$ when the promoter is unbound ($S = 0$) and $k_1$ when the promoter is bound ($S = 1$). Furthermore, a generalized rate function to describe binding/unbinding of a promoter

$$f(S, k) = k_- S + \frac{k_+ k_D}{k_M} k^2 (1 - S). \quad (7)$$

This function returns the unbinding rate $k_-$ when the promoter is occupied ($S = 1$) and the binding rate $\frac{k_+ k_D}{k_M} k^2$ if the promoter is free ($S = 0$). In this definition, $k$ is the molecular copy number of the TF that inhibits the promoter. With these definitions, we can formally write the CME for the toggle switch by listing all the reactions depicted in Figure 4B

$$\frac{dP(n, m, S_x, S_y, t)}{dt}$$
$$= -[\gamma n + \gamma m + k(S_x) + k(S_Y) + f(S_x, m) + f(S_y, n)]P(n, m, S_x, S_y, t) +$$
$$\gamma(n+1)P(n+1, m, S_x, S_y, t) + \gamma(m+1)P(n, m+1, S_x, S_y, t) +$$
$$k(S_x)P(n-1, m, S_x, S_y, t) + k(S_y)P(n, m-1, S_x, S_y, t)$$
$$+f(1-S_x, m)P(n, m, 1-S_x, S_y, t) - f(1-S_y, n)P(n, m, S_x, 1-S_y, t). \quad (8)$$

In eq. (8), the first-row accounts for all the outflux terms from the configuration $(n, m, S_x, S_y)$, including degradation, production and change in the promoters' states. The second-row accounts for influx due to degradation in configurations with higher molecule copy numbers; the third-row counts influx due to molecular production in configurations with lower copy numbers; and the fourth-row considers influx due to changes in the state of one of the promoters. This equation is far too complex for an exact solution with iterative methods; moreover, the relatively large number of variables makes eq. (8) stiff for generating function approaches. Simulations with the Gillespie algorithm, however, easily provide information about the relaxation dynamics and steady state (see for example [22,23]).

This treatment of the toggle switch can be considered as a footprint to tackle larger circuits of interconnected genes and transcription factors comprising transcription, translation, degradation, molecular binding and transcriptional regulation. As size of the circuits increases, understanding reasonable approximations to decrease the complexity of the model becomes crucial. Here, we applied some common approximations including QSSA for mRNA dynamics and protein-protein binding. In general, these approximations yield good results when timescales are well-separable. Therefore, depending on the parameters of the specific system of interest, QSSA might or might not be a suitable approximation.

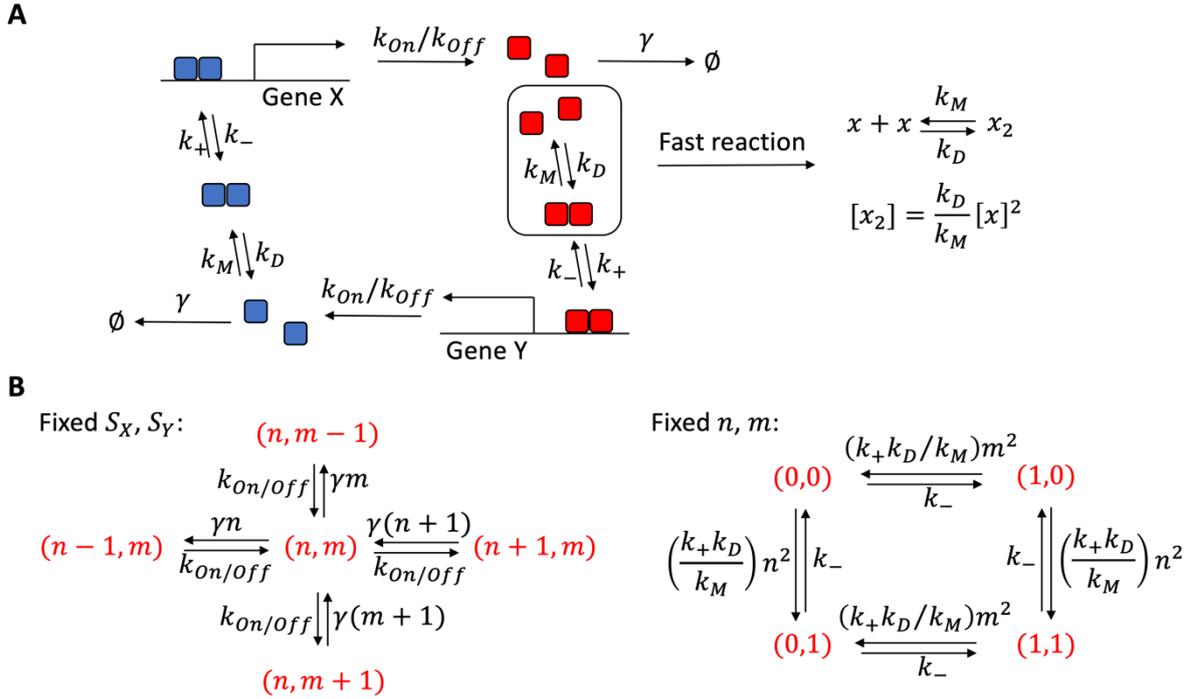

*Figure 4. **Circuit and reactions in a CME model of the toggle switch.** (A) In the toggle switch, gene x encodes for protein X (red blocks). X can dimerize and bind to DNA to inhibit the production of Y. Similarly, Y dimerizes and inhibits X. The dimerization reactions are fast compared to protein production and can thus be treated as instantaneously equilibrated. (B) Schematic of all the possible reactions leading to or out of the configuration $(n, m, S_x, S_y)$. For convenience, reactions are separated into two groups: changes in molecule copy number and changes in the state of activity of one of the promoters.*

---------------------------------------------------------------------------------------------------------------

2.1.4 Insight box 1: three ways to compute the steady state of the birth-death CME

In this insight box, we discuss three approaches to tackle the birth-death CME derived in section 2.1

$$\frac{dP_n}{dt} = -kP_n + kP_{n-1} - \gamma n P_n + \gamma(n+1)P_{n+1}. \quad (9)$$

Iterative solution of the CME

This infinite system of equations can be solved by taking advantage of the simplified form of the equation for $P_0$ and then using induction. Since the copy number is strictly positive ($n \geq 0$)

$$\frac{dP_0}{dt} = -kP_0 + \gamma P_1. \quad (10)$$

Steady state $dP_0/dt = 0$ implies $P_1 = (k/\gamma)P_0$. Similarly, $dP_1/dt = 0$ yields

$$P_2 = \frac{1}{2\gamma}(kP_1 - kP_0 + \gamma P_1) = \frac{1}{2}\left(\frac{k}{\gamma}\right)^2 P_0. \quad (11)$$

Repeating the process iteratively to higher-order equations, one finds the general expression for $P_n$

$$P_n = \frac{1}{n!}\left(\frac{k}{\gamma}\right)^n P_0, \quad (12)$$

Where the constant $P_0$ can be calculated by imposing normalization of the total probability

$$\sum_{n=0}^{\infty} P_n = P_0 \sum_{n=0}^{\infty} \frac{1}{n!}\left(\frac{k}{\gamma}\right)^n = P_0 e^{\frac{k}{\gamma}} = 1. \quad (13)$$

Thus, the steady state set of probabilities follows a Poisson distribution with rate $\lambda = k/\gamma$

$$P_n = \frac{1}{n!}\left(\frac{k}{\gamma}\right)^n e^{-\frac{k}{\gamma}}. \quad (14)$$

Generating functions

Generating functions advantageously transform an infinite-dimensional problem into a one-dimensional problem. Given the set of probabilities $\{P_n, n \propto [0, \infty]\}$, the generating function $\phi(z)$ is defined as

$$\phi(z) = \sum_{n=0}^{\infty} P_n z^n. \quad (15)$$

From eq. (15), an equation for the temporal dynamics of $\phi(z)$ can be obtained by taking a time derivative and substituting the actual expression of $dP_n/dt$. Then, the probabilities $P_n$ are obtained by differentiation $\phi(z)$

$$P_n = \frac{1}{n!}\frac{\partial^n \phi}{\partial z^n}(z = 0). \quad (16)$$

Using the definition of $\phi(z)$ of eq. (15), we write the single ODE that describes the temporal variation of $\phi(z)$

$$\frac{d\phi(z)}{dt} = \sum_{n=0}^{\infty} \frac{dP_n}{dt} z^n$$

$$= -k \sum_{n=0}^{\infty} P_n z^n + k \sum_{n=0}^{\infty} P_{n-1} z^n - \gamma \sum_{n=0}^{\infty} n P_n z^n + \gamma \sum_{n=0}^{\infty} (n+1) P_{n+1} z^n. \quad (17)$$

This equation can be tackled as follows. The first term on the RHS is simply $-k\phi(z)$. In the second term, we shift indexing to $m = n - 1$; then, it becomes apparent that the term equals $+kz\phi(z)$. In the third term, we can eliminate the factor $n$ by exploiting $\frac{d}{dz}(z^n) = nz^{n-1}$; then, the term can be rewritten as $-\gamma z \frac{d\phi(z)}{dz}$. Finally, combining the tricks used for the second and third terms, the fourth term becomes $\gamma \frac{d\phi(z)}{dz}$. The long-time limit solution $\frac{d\phi(z)}{dt} = 0$ is then

$$(1-z)\frac{d\phi(z)}{dz} = (1-z)\left(\frac{k}{\gamma}\right)\phi(z). \quad (18)$$

In any point $z \neq 1$, this equation has the simple general solution $\phi(z) = C_1 e^{\frac{k}{\gamma}z} + C_2$. The probabilities $\{P_n, n \propto [0, \infty]\}$ are then obtained by differentiating the generating function

$$P_n = \frac{\phi^n(0)}{n!} = \frac{1}{n!}\left(\frac{k}{\gamma}\right)^n C_1. \quad (19)$$

This is the same expression found with the iterative method, with the parameter $C_1$ in place of $P_0$. Therefore, normalization of the probability set $\{P_n\}$ yields $P_n = \frac{1}{n!}\left(\frac{k}{\gamma}\right)^n e^{-\frac{k}{\gamma}}$.

Gillespie simulation

In the simple case of the birth-death process, starting from a generic configuration at time $t$ with $n$ molecules, two reactions are possible: birth of a new protein molecule with rate $k$ and death of an existing molecule with rate $\gamma n$. Below we provide the pseudocode for the implementation of the Gillespie scheme for the birth-death process.

Pseudocode:

$n = n_{in}$  # set the initial number of molecules
$t = 0$     # start the simulation at t=0

```
while t < T:
    # compute total rate
    W = k + γn
    # sample an exponentially-distributed random variable with rate parameter 1/W
    τ = exp_rnd(1/W)
    # sample a uniformly distributed random number in [0,1] to pick the reaction
    # select birth reaction
    if uniform([0,1]) < k/W:
        n = n + 1
    # select death reaction
    else:
        n = n - 1
    # update time
    t = t + τ
```

In the birth-death process case, the birth rate is configuration-independent and always equal to $k$. The death rate, however, depends on protein number and is minimal at $n = 1$. Therefore, the simulation time ($T$) required to obtain substantial information about the steady state must at least satisfy $T \gg \max(1/k, 1/\gamma)$.

---------------------------------------------------------------------------------------------------------------------

2.1.5 Explore further: RNA splicing

In this section, we mostly focused on transcription (production of mRNA) and translation (production of the finalized molecule) as the building blocks of cellular reaction systems. Newly transcribed RNA molecules, however, require a specific reaction where pre-determined portions of the RNA sequence are removed. This "splicing" reaction converts nascent to mature RNA, which can be then effectively translated into protein [24]. mRNA splicing has recently drawn interest in the systems biology field due to the development of new single cell sequencing technologies that now enable to precisely measure the amount of unspliced (nascent) and spliced (mature) RNA within individual cells [25,26]. This high-dimensional information has been used to build biochemical reaction models that infer interactions between chemical species and predict transitions between cellular states [27–29]. A detailed discussion of the mathematical approaches and consequences of splicing in the molecular dynamics can be found in [30,31].

**2.2 Continuous models of regulatory networks**

2.2.1 The chemical rate equation emerges as the average dynamics of the CME

In the previous section, we developed a probabilistic framework to account for the inherent stochasticity of biochemical reactions in regulatory networks. This probabilistic framework is especially important when the copy number of a protein is low within a cell, and thus stochastic fluctuations can substantially influence a biochemical circuit's response and potentially play a role in cellular function. A well-known example of the role of fluctuations is the selection of a competent state for DNA uptake in *Bacillus subtilis* [32]. In the limit of large molecular copy number, however, one can assume that stochastic fluctuations become less and less relevant (also referred to as "thermodynamic limit"). In this limit, the concentration or copy number can be described as a continuous variable. In more practical terms, when moving from the CME to a continuous formalism, we coarse-grain the model of a biochemical network by describing each chemical species with a single, continuous variable, whose temporal dynamics can be described by a single ordinary differential equation (ODE), rather than an infinite-dimensional system of CMEs.

The relation between the CME and the continuous chemical rate equation can be understood in the simple case of a birth-death process presented in eq. 4 by considering the dynamics of the average number of proteins $\langle n(t) \rangle$

$$\frac{d\langle n(t)\rangle}{dt} = \sum_{n=0}^{+\infty} n \frac{d\langle P_n(t)\rangle}{dt}$$

$$= k \sum_{n=0}^{+\infty}[nP_{n-1}(t) - nP_n(t)] - \gamma \sum_{n=0}^{+\infty}[n^2 P_n(t) - n(n+1)P_{n+1}(t)]. \quad (20)$$

After some manipulation of the summation indexes, eq. (20) can be simplified as

$$\frac{d\langle n(t)\rangle}{dt} = k - \gamma\langle n(t)\rangle. \quad (21)$$

Therefore, the evolution of the average number of molecules, which is now a continuous variable, is described by a single ODE.

This approach facilitates the treatment of large networks with many interconnected regulations that would be unfeasible with the CME approach. In general, a circuit of N interconnected species can be described as a dynamical system

$$\dot{\boldsymbol{x}} = F(\boldsymbol{x}), \quad (22)$$

where $\boldsymbol{x} = (x_1, x_2, \ldots, x_N)$ is the vector of concentration/copy number of the N species in the circuit and the force field $F(\boldsymbol{x})$ describes their coupled dynamics and the possible interconnections between them. Therefore, all the prior knowledge on the circuit of interest is encoded by the choice of functions and parameters in the force field. The force field $F(\boldsymbol{x})$ captures all the relevant reactions, including transcription, translation, chemical binding and degradation. The continuous formulation of eq. (22) conveniently allows to apply all the tools of nonlinear dynamics, including linear stability analysis, phase diagrams and bifurcations that are usually applied to classical physical problems. In the following section, we will develop some of these methods in the study of the continuous toggle switch.

Finally, it is worth noting that significant deviations between CME and mean field modeling may arise when mean field models are pushed to the mesoscale where stochastic fluctuations become important. The presence of different states, or cell phenotypes, is represented in the CME formalism by multimodal probability distributions whereby each peak corresponds to a cell phenotype, as seen in the self-activating gene (Fig. 3). In the framework of deterministic ODE models, the accessible phenotypes are represented by the stable attractors. The correspondence between distribution peaks and stable attractors is generally good in the thermodynamic limit but discrepancies can arise in the low copy number limit (see [33,34] for detailed comparisons). In addition, deterministic models cannot capture the transitions between cell states guided by stochastic fluctuations. The effect of noise-driven phenotype switching can be introduced in ODE-based modeling with the stochastic differential equation framework, which will be introduced in section 2.2.3. More detailed mathematical insight and a thorough comparison between the stochastic and mean field approaches can be found in [35].

2.2.2 Continuous model of the toggle switch

Following the parallel with the treatment of the CME, we develop a continuous formalism for the toggle switch system composed of two genes that mutually repress each other. We also exploit this example to present some common functions used to model transcriptional activation and inhibition in the gene regulatory network literature. In the CME, the mutual inhibition between two TFs was explicitly modeled by considering TF dimerization and binding to the DNA. The main challenge of the continuous version is to find a suitable force field $F(x)$ that effectively captures the same dynamics. This is typically achieved by introducing a continuous function that modulates the transcription rate. First, we can introduce transcriptional regulation in the simpler case of the noise-free chemical rate equation of the birth-death process

$$\frac{dp}{dt} = k\, H(R) - \gamma p. \quad (23)$$

Here, $R$ is the concentration of a transcription factor that inhibits or activates the production of $p$; $H(R)$ is a continuous function that modulates the production rate. In general, $H(R)$ must satisfy the condition $H(0) = 1$ so that the basal transcription rate $k$ is recovered in the absence of external regulation. Then, $H(R)$ increases or decreases monotonically as a function of $R$ depending on whether $R$ activates or inhibits $p$. Assuming equilibration of both TF dimerization/multimerization and TF-DNA binding leads to the so-called Hill function, which was originally introduced in 1910 to describe the binding of oxygen to hemoglobin [36], and has been since applied to a variety of biological contexts [37]

$$H(R) = \frac{1 + \lambda \left(\frac{R}{R_0}\right)^n}{1 + \left(\frac{R}{R_0}\right)^n}. \quad (24)$$

The functional form of the Hill function in Eq. (24) is derived exactly from the underlying chemical reactions in the insight box 2. In this expression, $R_0$ represents a threshold concentration that is related to the rate constants for TF-TF and TF-DNA binding (see insight box 2). Therefore, when $R > R_0$ the regulatory effect of $R$ becomes important. $\lambda$ is a fold-change that represents the change in transcription rate due to $R$. In the limit $R \gg R_0$, the production rate relaxes to $k\lambda$. Therefore, $0 < \lambda < 1$ indicates transcriptional inhibition and $\lambda > 1$ indicates transcriptional activation. Lastly, the Hill coefficient $n$ indicates the cooperativity of the transcription factor. For instance, $n = 1$ indicates that $R$ binds to the promoter of $p$ as a single molecule, $n = 2$ indicates dimerization of the TF, and so forth. It is important to stress that a Hill-like relation between protein and TF does not necessarily imply TF cooperation and binding, as other biological processes upstream or downstream of transcriptional regulation that are not accounted for in the model can lead this nonlinear response of eq. (24). The insight box 2 (section 2.2.5) further discusses how these parameters can be fitted or inferred from experiment observations.

In this continuous formalism, the toggle switch can be thus described by a set of two ODEs of the form

$$\frac{dx}{dt} = k \frac{1 + \lambda \left(\frac{y}{t_0}\right)^n}{1 + \left(\frac{y}{t_0}\right)^n} - \gamma x, \quad (25a)$$

$$\frac{dy}{dt} = k \frac{1 + \lambda \left(\frac{x}{t_0}\right)^n}{1 + \left(\frac{x}{t_0}\right)^n} - \gamma y. \quad (25b)$$

For simplicity, we assumed that both $x$ and $y$ are produced and degraded with the same rate constants. Furthermore, the system is completely symmetric as the Hill thresholds, fold-changes and coefficients are the same for the two species. The ODE system of eq. (25) can exhibit bistability between a (high X, low Y) and a (low X, high Y) states (Fig. 5A). A potential pitfall of this modeling approach is the uncertainty in the estimation of the model's parameters. The dependence on the parameter choice can be investigated with parameter sensitivity approaches. Figure 5B shows the change of steady state X level in the (high X, low Y) state upon varying the model's parameter one at a time by 5%. More sophisticated approaches to parameter sensitivity exist in the current literature that can be reviewed in references[38,39].

While the toggle switch represents a simple case of bistability of two opposite states, often the description of realistic biological processes requires more complex models, which can be achieved by increasing the number of nodes and interactions. For example, allowing both species to self-activate enables a third stable state with intermediate expression of both $x$ and $y$. This self-activating toggle switch can describe the differentiation of an undecided stem-like state into either one of two differentiate states that either express $x$ or $y$ [40], and will be further discussed in the context of stem cell differentiation (see section 3.2). Recently, a systematic network motif search by Ye and collaborators identified two large families of 3-node networks that give rise to 4 stable cell states. In this study, the authors first sampled different network topologies defined as the sets of positive and negative interactions between nodes, and then developed a mathematical for each topology to determine the number of attractors. Finally, these multi-stable circuits were used to describe the differentiation of T-cells, where an undifferentiated progenitor transitions through several intermediate states before reaching a differentiated state[41].

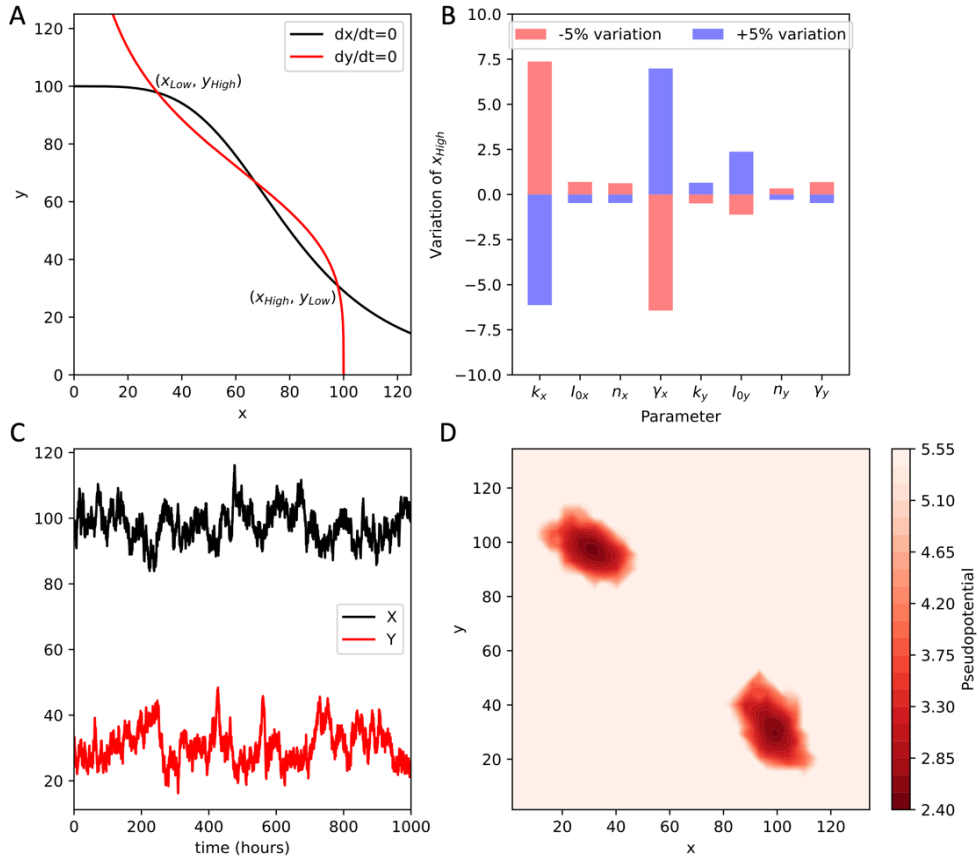

*Figure 5. The deterministic and stochastic toggle switch. (A) The phase space of the toggle switch described in eq. (25) including nullclines and fixed points. (B) Local sensitivity analysis of the toggle switch parameter around the (high X, low Y) stable fixed point. (C) A temporal trajectory for the stochastic toggle switch. (D) The pseudopotential landscape of the toggle switch. The white noise level is fixed to $\sigma = 2$ in panels C, D. The source code to reproduce these results is available in our [toggle switch tutorial](#).*

### 2.2.3 Introducing noise in continuous circuits with the overdamped Langevin equation

We derived the continuous chemical rate equation as a noise-free approximation to the CME. In many situations, however, the continuous limit is utilized because the complexity and size of the circuit of interest prevents a treatment with discrete modeling. The effect of stochastic fluctuations, however, can be still incorporated in continuous models in a coarse-grained manner by considering the stochastic differential equation

$$\dot{x} = F(x) + \xi(x, t). \quad (26)$$

This equation describes the motion of a particle in the overdamped limit ($\ddot{x} = 0$) under the presence of a force $F(x)$ and a noise term $\xi(x, t)$ that satisfies $\langle \xi(x, t)\xi(x, t')\rangle = 2D\, \mathbf{D}\, \delta(t - t')$. In other words, the noise term is a random variable whose intensity at any time $t$ is completely uncorrelated to the intensity at any other given time. $D$ is a constant with the interpretation of an effective diffusion coefficient. In most of the current models of genetic circuits, the noise is assumed to be white. Thus, $\mathbf{D}$ is simply a unitary matrix. In other words, the noise term for each species has the form $\xi(x, y, t) = \sigma\, N(0,1)$, where $\sigma$ is the noise amplitude and $N(0,1)$ is a Gaussian random variable with zero mean and unitary variance. In the case of the bistable toggle switch, white noise introduces stochastic fluctuations can induce transitions between the two stable states (Fig. 5C).

It is important to stress that stochastic fluctuations modeled in the previous section with Gillespie simulations depend on the models' parameters. In other words, each chemical species had a distinct noise amplitude, which could furthermore depend on time as the different reaction rates may depend on many variables' copy numbers. For this reason, the stochastic fluctuations arising from biochemical reactions and modeled with Gillespie-style simulations are referred to as 'intrinsic noise'. Conversely, the so-called 'extrinsic noise' broadly captures all other perturbations related to other cellular components and/or external factors in the cell's local microenvironment [42,43]. Modeling fluctuations with white noise implies that all species in a circuit are subjected to a fixed noise intensity. From a statistical perspective, the approximation of stochastic fluctuations with white noise can be motivated, in the thermodynamic limit, by the central limit theorem. For example, the protein distribution in the birth-death process is well-approximated by a Gaussian fit (see again Fig. 2A). On a more physical basis, white noise emerges, under specific conditions, whenever a system receives and dissipated energy from a "bath" (in this case, the rest of the cell) that is not explicitly included in the model [44,45]. This assumption, however, could become problematic when modeling species that are lowly expressed or near "extinction" points where all molecule copies are degraded. Protein copy numbers vary significantly across species and specific conditions. Copy numbers in human cells span as much as 7 orders of magnitude, and low-expressed proteins can have less than 500 copies, whereas transcription factors in bacteria can have only tens of copies [46,47]. For example, the behavior of the $\lambda$-phage virus has been described using a toggle switch model between the repressors *cI* and *cro* whereby high cI implies lysogeny where the virus is incorporated into the host's DNA while high *cro* implies lysis where the virus reproduces and kills the host cell [48]. Detailed mathematical analysis by Schultz and colleagues showed that a stochastic ODE model cannot correctly account for the bistability as *cro* can undergo "extinction" and "resurrection" events, where *cro* is temporarily absent before being produced again [49].

To bridge the gap at least partially between continuous and discrete model at low molecule copy numbers, other ways to model noise have been explored. One notable example is shot noise where $\boldsymbol{D}$ is a diagonal matrix with $D_{ii} = x_i$ [44,45]. In other words, the intensity of noise for the $i$-th species is proportional to the square root of the variable's copy number ($\sqrt{x_i}$). Therefore, variables with higher copy number experience fluctuations with larger amplitude. On the other hand, the noise-to-signal ratio $\sqrt{x_i}/x_i = 1/x_i$ is lower for high copy number species. As a practical example, a species with average copy number $n = 10^2$ molecules will have typical fluctuations of ~10 molecules, or 10% of its average copy number. Conversely, the average fluctuations of a species with copy number of $n = 10^4$ will be larger in absolute intensity (~$10^2$ molecules) but will only correspond to 1% of the variable's copy number. Thus, species with low copy number can be subjected to large deviations somewhat similar to the bursting effect discussed in the context of the CME. As noted by Lu and collaborators, a combination of white and shot noise is likely a more precise way to model stochastic fluctuations in continuous, ODE model of biochemical regulatory networks as shot noise captures low copy number deviations while white noise captures external inputs [50].

### 2.2.4 Characterizing the landscape of multistable gene regulatory networks

In general, a multistable circuit governed by the overdamped Langevin equation will exhibit multiple attractors and, as in the case of the bistable toggle switch, stochastic fluctuations introduce the possibility of transitions from one state to another. Landscape theory is a popular way to quantify the global stability in a multistable system by computing the equivalent of a potential function.

The idea that cells follow "paths" in some underlying, high-dimensional landscape that guide the transition toward new cell states was first formalized by Waddington. In Waddington's epigenetic landscape, cells are visualized as marbles that roll through a downhill landscape where valleys and ridges define alternative routes leading to different cellular phenotypes characterized by distinct gene expression profiles and different sets of epigenetic interactions [51]. Here, the term "epigenetics" denotes the set of intracellular interactions between the genome, RNAs, transcription factors and enzymes, such as DNA methylation and phosphorylation, that modulate gene expression. Therefore, the differences in gene expression and protein concentration in cells belonging to different valleys is not a result of mutations in the genetic sequence, and rather arise from different epigenetic interactions operating on the same genome [52,53].

The construction of the landscape begins by mapping the Langevin dynamics onto a Fokker-Planck (FP) equation describing the joint probability distribution $p(x_1, x_2, \ldots x_N, t)$

$$\frac{\partial p(\boldsymbol{x},t)}{\partial t} = -\nabla \cdot [F(\boldsymbol{x})p(\boldsymbol{x},t)] + \nabla^2[D(\boldsymbol{x})p(\boldsymbol{x},t)], \quad (27)$$

where $F(\boldsymbol{x})$ is the force field driving the Langevin dynamics and $D(\boldsymbol{x})$ is a diffusion tensor related to the noise amplitude by $D(\boldsymbol{x}) = \sigma(X)\sigma^t(x)/2$. We will not discuss the derivation of the FP equation form the Langevin equation, which can be found in many textbooks on non-equilibrium statistical mechanics. The FP equation can be written more compactly in the form of a continuity equation

$$\partial p(\boldsymbol{x},t)/\partial t + \nabla \cdot \boldsymbol{J} = 0, \quad (28)$$

where $J(\boldsymbol{x},t) = F(\boldsymbol{x})p(\boldsymbol{x},t) - \nabla \cdot [D(\boldsymbol{x})p(\boldsymbol{x},t)]$. From eq. (28), it is evident that the steady state solution $\partial p(\boldsymbol{x},t)/\partial t = 0$ requires a divergence-free flux $\nabla \cdot \boldsymbol{J} = 0$. This condition can be satisfied in different ways. Systems that respect detailed balance satisfy the stringent condition $J = 0$; in this case, there is a clear relation between force, potential and equilibrium probability distribution. Indeed, enforcing $J = 0$ implies

$$F(x) = \frac{\nabla \cdot [D(x) p_{eq}]}{p_{eq}} = D(x) \frac{\nabla P_{SS}}{P_{SS}} + \nabla \cdot D(x). \quad (29)$$

Rewriting $\frac{\nabla P_{eq}}{P_{eq}} = -\nabla \log(P_{eq})$ clarifies the relation between potential and the equilibrium probability distribution $U = -\log(P_{eq})$. The dynamics of the system is completely determined by the gradient of the potential $U$ (up to a shift due to local variation of the diffusion matrix $\nabla \cdot D(x)$). In the simple case of white noise, the diffusion is uniform ($\nabla \cdot D(x) = 0$) and the correspondence between force and potential gradient is exact. Typically, force fields describing regulatory networks are constructed in a phenomenological manner to capture biochemical interactions such as transcription, binding or degradation with the use of functional forms such as Hill functions. For this reason, they cannot be derived from an underlying potential and do not satisfy $J = 0$. Therefore, at steady state, the force field can be written as the sum of two components. A gradient of the log-scaled steady-state probability distribution and flux term

$$F(x) = D(x) \cdot \frac{\nabla P_{SS}}{P_{SS}} + \nabla D(x) + \frac{J_{SS}}{P_{SS}}. \quad (30)$$

The flux term $J_{SS}$ effectively quantifies the deviation from gradient-driven dynamics. Since there are no sources or sinks of probability in the domain of existence of $x$, $J_{SS}$ must be locally curled, and is therefore often referred to as the curl flux. As observed by Wang, the dynamics of 'gradient' dynamical systems that respect detailed balance can be compared to the motion of an electron in an electric field; conversely, the dynamics of 'non-gradient' systems such as gene regulatory networks is similar to the motion of an electron in an electromagnetic field [54].

Fig. 5D shows the pseudopotential landscape of the bistable toggle switch under the effect of white noise, which include two deep minima corresponding to the two stable fixed points of the deterministic system.

Therefore, in non-gradient systems such as coupled biochemical networks, the steady-state probability distribution $p(x, t)$ cannot be established *a priori*, but only from stochastic simulations of the Langevin equation or steady state solution of the FP equation.

Nonetheless, the steady-state probability distribution and curl flux provide key information about (1) the number of accessible states, (2) their relative stability, and (3) the deviation from gradient-driven dynamics.

As seen in the self-activating gene (Figure 3) and the toggle switch (Figure 5), noise can induce cell state transitions between attractors. While these transitions can be studied with long simulations of the overdamped Langevin equation (such as the trajectory in Fig. 5C), this approach is time consuming and often cannot provide large statistics because escape events from attractors are rare. The pseudopotential landscape offers a more general strategy to study transitions with methods based on the path integral formalism. These methods have been widely applied to biochemical systems to reconstruct transition trajectories and compute transition rates in a more computationally efficient manner. Interesting examples of these approaches include [55,56], and are summarized in [57] for interested readers.

---

2.2.5 Insight box 2: Modeling transcriptional regulation with Hill functions

In the simplest model of protein production, a gene transcribes mRNA at a constant rate $k_m$; in turn, mRNA molecules are translated into proteins with rate constant $k_p$. Therefore, the copy number of mRNA (m) and protein (p) can be described by first order equations

$$\frac{dm}{dt} = k_m D_0 - \gamma_m m, \quad (31a)$$

$$\frac{dp}{dt} = k_p m - \gamma_p p, \quad (31b)$$

where $D_0$ is the total number of promoters transcribing for the protein p, and can be treated as a constant unless DNA is being replicated during cell division; $\gamma_m$ and $\gamma_p$ are degradation rate constants for mRNA and protein. Assuming fast equilibration of mRNA ($dm/dt = 0$), the mRNA copy number is $m = k_m D_0 / \gamma_m$ and the protein copy number directly depends on the number of genes $D_0$

$$\frac{dp}{dt} = kD_0 - \gamma_p p, \quad (32)$$

where $k = k_m k_p/\gamma_m$. In presence of a transcriptional factor $x$ that activates or inhibits gene expression, the mRNA production rate is modulated by a function $k_m \to k_m f(x)$ (Figure A). First, we assume that the transcription factor can bind and unbind to DNA as a single molecule with rate constants $k_+, k_-$. Assuming fast equilibration of the TF-DNA binding, the fraction of genes that are unbound ($U$) and bound ($B$) to the transcription factor are

$$U = D_0 \frac{1}{1 + \frac{x}{x_0}}, \quad (33a)$$

$$B = D_0 - U = D_0 \frac{\frac{x}{x_0}}{1 + \frac{x}{x_0}}, \quad (33b)$$

where $x_0 = k_+/k_-$ represents the half-concentration so that $U = B$ when $x = x_0$. The mRNA is transcribed with rate constant $k_u$ by an unbound promoter and $k_b$ by a bound promoter, respectively. $k_b > k_u$ implies that $x$ is an activator while $k_b < k_u$ implies that $x$ is an inhibitor. The corresponding protein ($p$) is then produced with rate $(k_u k_p/\gamma_m)U + (k_b k_p/\gamma_m)B$. Thus, the dynamics of protein $p$ is governed by the following equation

$$\frac{dp}{dt} = k^p{}_U \frac{1}{1+\frac{x}{x_0}} + k^p{}_B \frac{\frac{x}{x_0}}{1+\frac{x}{x_0}} - \gamma_p p = k^p{}_U \frac{1+\lambda\frac{x}{x_0}}{1+\frac{x}{x_0}} - \gamma_p p, \quad (34)$$

where $k^p{}_U = (k_u k_p/\gamma_m)$ and $k^p{}_B = (k_b k_p/\gamma_m)$. In the compact form on the right-hand side, $\lambda = k^p{}_B/k^p{}_U$ represents a fold-change in the transcription of $p$ due to the regulatory activity of $x$.

In the more complex case where the transcription factor dimerizes reversibly ($x + x \leftrightarrow x_2$) with dimerization and monomerization rate constants $k_2, k_1$, and binds and unbind to DNA only in its dimer form $x_2$, the fractions of bound and unbound DNA become

$$U = D_0 \frac{1}{1 + \left(\frac{x}{x_0}\right)^2}, \quad (35a)$$

$$B = D_0 - U = D_0 \frac{\left(\frac{x}{x_0}\right)^2}{1 + \left(\frac{x}{x_0}\right)^2}, \quad (35b)$$

where $x_0 = \sqrt{k_1 k_+/k_2 k_-}$. The corresponding equation for protein $p$ becomes:

$$\frac{dp}{dt} = k^p{}_U \frac{1 + \lambda \left(\frac{x}{x_0}\right)^2}{1 + \left(\frac{x}{x_0}\right)^2} - \gamma_p p. \quad (36)$$

In general, a transcription factor that binds to DNA as an n-oligomer is described by the n-th power of its monomeric concentration. If the transcription factor is an activator, it is often assumed that the transcription rate of the unbound gene is zero (i.e. $k_u = 0$), hence

$$\frac{dp}{dt} = k \frac{\left(\frac{x}{x_0}\right)^n}{1 + \left(\frac{x}{x_0}\right)^n} - \gamma_p p, \quad (37)$$

where the concentration of DNA ($D_0$) has been absorbed into the production rate constant $k$. Similarly, if a transcription factor is a very effective inhibitor, no transcription is assumed for the bound genes (i.e. $k_b = 0$), hence

$$\frac{dp}{dt} = k \frac{1}{1 + \left(\frac{x}{x_0}\right)^n} - \gamma_p p. \quad (38)$$

The expressions in eq. (37) and eq. (38) are referred to as positive and negative Hill functions, whereas the general form presented in eq. (36) is referred to as shifted Hill function. Detailed data about TF oligomerization is not always available. The Hill coefficient, half-concentration and fold-change ($n$, $x_0$, $\lambda$), however, can still be inferred from an experiment describing protein concentration as a function of TF concentration (Figure 6B). In this case, however, there is no guarantee that the inferred Hill coefficient truly reflects TF oligomerization, as other processes can modulate transcription and change the protein-TF relation.

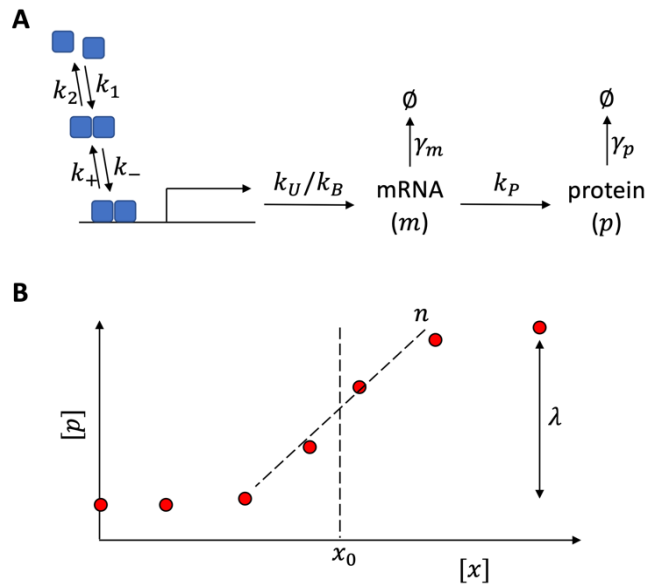

*Figure 6. **Derivation and fitting of the Hill function.** (A) The derivation of the Hill function considers a case where the regulator can dimerize (or create a higher-order oligomer) and then bind to DNA. (B) The parameters of the Hill functions can be inferred from experimental data reporting the expression of the molecule $p$ as a function of transcription factor concentration.*

-------------------------------------------------------------------------------------------------------

2.2.6 Explore further: Time delay, molecular memory and noncanonical pathways to multistability

A key assumption of both the CME and Langevin modeling strategies is that all biochemical reactions are Markovian with a Poisson waiting time distribution. Several biochemical processes, however, do not satisfy these assumptions. Protein synthesis involves the sequential assembly of long molecules, wherein a multistep process is aggregated into a simpler, single-step reaction described by a single rate constant, thus effectively implying a minimal waiting time for transcription or translation to occur[58,59]. The idea that the synthesis or degradation of a macromolecule involves a cascade of multiple, sequential events is referred to as "molecular memory"[60]. Non-Markovian dynamics has been modeled in different ways, often leading to diverse mathematical behavior that could not be obtained under Markovian assumptions. The auto-regulatory motif where a transcription factor (TF) inhibits its own transcription is a clear example that illustrates this concept. Experimentally, this motif can lead to oscillations in TF concentration, as the time required for TF synthesis and diffusion inside the cell nucleus introduces a temporal delay. This effect can be modeled by introducing a delay in the transcription rate: $f(TF(t)) \to f(TF(t-\tau))$. Thus, the transcription at time $t$ is a function of TF concentration at time $t - \tau$, where the specific values of delay and degradation rate parameters determines the period of the oscillations[61]. This modeling strategy has been applied to many models involving transcription factors such as p53[62], NF-kB[63], and PAGE4[64], yielding a good description of the oscillatory behavior that could not be captured by a "standard" Markovian model including instantaneous transcription, translation, and repression. Moreover, spatial models of partial differential equations have been developed to describe in greater details the timescale arising from mRNA and protein diffusion within cells, thus avoiding the need to explicitly introduce delay terms in the model's parameters[65]. Recent systematic studied of the mathematical conditions enabling oscillations in in gene regulatory networks can be found in[66,67].

Besides non-Markovian reactions, multistability can be achieved via "noncanonical" mechanisms that do not simply rely on transcriptional activation or inhibition. A first example is the action of microRNAs, specific RNA species that bind to messenger RNA to promote their degradation. Recently, a model of mRNA-microRNA interactions was used to explain the spatial segregation of transcription factors Hoxa5 and Hoxa8 during spinal cord development[68]. In this context, a toggle-switch type mutual inhibition between Hoxa5 and Hoxa8, although theoretically suitable to explain the bistability, was disproven by experimental evidence suggesting unilateral inhibition from Hoxa8 toward Hoxa5[68]. The detailed mathematical modeling of microRNA-assisted mRNA degradation will be discussed later (section 3.1.3) in the context of epithelial-mesenchymal transition. A second noncanonical pathway to bistability without transcriptional regulation involves phosphorylation by enzymes. The activation of Mitogen-activated protein kinases (MAPK), a class of proteins that convert and process extracellular stimuli, requires two sequential phosphorylation events carried by different enzymes at two sites, whereby a first enzyme phosphorylates the first site and then releases the intermediate kinase before a second phosphorylation event occurs. Markevich and collaborators demonstrated that the competitive binding of different enzymes to the MAPK binding sites is sufficient to generate bistability without additional transcriptional regulation[69].

## 2.3 Large networks: Boolean models, parameter randomization and interfacing with data

In the previous two sections we discussed CME and continuous models that describe genetic and biochemical interactions in a mechanism-based, detailed manner. These approaches can become impractical when applied to large networks with dozens, or even hundreds, of chemical species. These considerations are becoming more and more relevant as detailed experimental techniques provides detailed insight into the organization of large regulatory networks, thus providing an opportunity to build and benchmark larger models of gene regulation. A first challenge, from a computational standpoint, is to perform long simulations on models with hundreds of reactions and/or ODEs. A second, more fundamental problem is the difficulty to precisely estimate all the parameters of the model. Quite often, only a qualitative relation can be established between species in a network, such as inhibition or induction. In this section, we discuss two classes of models that resolve these issues and allow the modeling of large networks. First, we discuss Boolean models where nodes can only be active or inactive, and transcriptional and biochemical interactions are described via logic operators. Second, we present parameter randomization approaches that assume the functional form of the interactions and test many different combinations of parameter values. Finally, we introduce how the problem of determining the gene regulatory network architecture and connections can be tackled by interfacing quantitative modeling and high-resolution single cell sequencing data.

2.3.1 Boolean Models

Boolean models perhaps represent the most coarse-grained approach to describe regulatory networks since they require a minimal amount of information including solely the circuit's connections of positive or negative interactions among genes [70]. For this reason,

Boolean approaches stand out as some of the first attempts to model gene regulation [71]. In Boolean models, each node can either be ON ($\sigma = 1$) or OFF ($\sigma = 0$), thus representing a gene or molecular species that is active or inactive, respectively. Therefore, the state of a regulatory network of N genes at any given time $t$ is specified by an array of N Boolean variables $\sigma = (\sigma_1, \sigma_2, \ldots, \sigma_N)$. The value of each node $i$ ($\sigma_i$) is a function of all the incoming signal from nodes that directly regulate node $i$. These functions $B_i$ are typically constructed in an *ad hoc* manner starting from experimental evidence and using the logic commands AND, OR and NOT. Similar to their continuous counterparts, Boolean networks can be represented graphically as a set of nodes and arrows to indicate mutual activation or inhibition, such as shown in figure 7A. It is important to stress that the functional form of the functions $B_i$ cannot be inferred by simply looking at the schematic representation. For example, in the circuit of figure 7A, x and z both activate the node y. Their input, however, can be combined in an AND or OR manner, which must be specified.

Starting from an initial configuration, the values of the nodes are updated according to an integration scheme. In particular, the synchronous scheme is the simplest update method on a Boolean network. Under this deterministic scheme, the state of each node ($i$) at iteration $t + 1$ is the output of the Boolean function $B_i$ computed at time $t$

$$\sigma_i(t+1) = B_i(\boldsymbol{\sigma}(t)). \quad (39)$$

This method presents some important drawbacks. Specifically, reactions in regulatory networks can be time-separated, with given reactions evolving on faster timescales than others. A popular solution is given by asynchronous updating schemes, which can be considered as a Boolean equivalent of the Gillespie algorithm. With this method, any reaction in the network is associated with a typical timescale. At any updating step, one node is selected with a probability proportional to the frequency of that reaction. Therefore, faster reactions are more probable and thus selected more often. These reference timescales are not known *a priori*, so they must be inferred from literature and given as an additional input in the construction of the model. A more in-depth discussion and comparison between different integration schemes is provided in [72,73].

Starting from a given initial condition, a Boolean network relaxes to an attractor or limit cycle by following the update scheme. In figure 7A, we present a simple example of a 3-nodes network originally proposed by Wang and collaborators [74] with the following Boolean functions

$$B_x = x \; OR \; (NOT \; z), \quad (40a)$$

$$B_y = x \; AND \; z, \quad (40b)$$

$$B_z = y. \quad (40c)$$

Three different initial conditions lead to three different steady states, including two stable attractors and a limit cycle (Figure 7B). More generally, a large Boolean network with N nodes has $2^N$ different initial conditions that can lead to multiple attractors and/or limit cycles. Typically, their basin of attraction is quantified with a large number of simulations starting from randomized initial conditions.

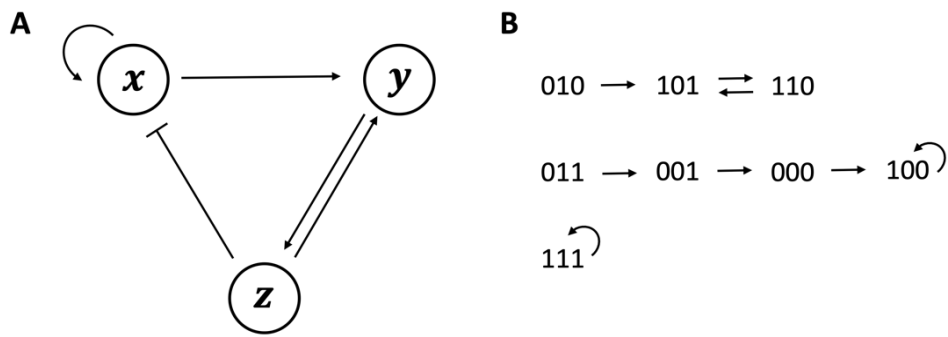

*Figure 7. (A) Example of a three-node Boolean model. (B) Three different initial conditions that lead to oscillations between two configurations*

2.3.2 Parameter randomization

Boolean models overcome the insufficient knowledge of the model's parameters by assuming more coarse-grained relationships between species based on logical operators, thus sacrificing the detailed description of biochemical reactions. An alternative approach is to hypothesize more detailed functional forms to describe the circuit's reactions, such as hill functions for transcriptional regulation, and explore several parameter combinations.

For example, _ra_ndom _ci_rcuit _pe_rturbation (RACIPE) takes the topological information of a network (i.e., the list of positive and negative interactions between species) as the input and generates an ensemble of mathematical models[75]. Each mathematical model, also referred to as a RACIPE model, is simulated with the same set of chemical rate equations but with distinct sets of parameters. Each set of parameters are randomly sampled from a fixed distribution (uniform, Gaussian etc.) within their biologically reasonable ranges. For each RACIPE model, multiple (typically tens of thousands) initial conditions are used to perform the numerical simulation to solve all possible stable-state solutions. The stable-state solutions from the ensemble RACIPE models and their corresponding parameters are collected for statistical analysis, by which the pattern of solutions and the differences in parameters leading to different solutions can be identified. RACIPE originally focuses on transcriptional regulatory networks and uses the shifted Hill function[76] to simulate the transcriptional regulations. The method RACIPE has been implemented as a free open source software which can be accessed in GitHub[77]. To reach convergence of RACIPE simulation results, there are two key parameters - the number of RACIPE models and the number of initial conditions for each model should be evaluated and decided. As the average simulation time of a RACIPE model is linearly proportional to the number of parameters of that model, RACIPE can be potentially applied to large gene networks. Notably, the main purpose of RACIPE is to determine the robust stable states enabled by the gene networks by considering the large variations in parameters. For example, parameter randomization of the toggle switch topology using RACIPE showed that, even though the toggle switch topology is typically associated with bistability, more than 75% of parameter combinations lead to a single stable fixed point[75].

To detect the stability of different stable states exhibited by the gene network, a method integrating RACIPE with stochastic analysis, referred to as sRACIPE, has been recently developed[78]. The sRACIPE employs a simulated annealing-based scheme to estimate the stability of different stable states resulting from various initial conditions. One interesting observation by applying sRACIPE to toggle-switch-like circuits is that high expression noise induced state merging. Recently, the Boolean and randomized approaches were compared on the same set of network topologies, showing that both methods can successfully recover the important dynamical features of the circuits[79].

2.3.3 Learning regulatory networks from single cell transcriptomics

So far, we have always assumed the existence of biological knowledge that serves as input to build GRN models. In many cases, however, the interactions between genes are not well known, and alternative, data-driven strategies are necessary. Over the last decade, many single cell Omics technologies have started to provide unprecedented resolution over gene regulation at the single cell level [80]. In this section, we focus specifically on single cell RNA sequencing (scRNA-seq), a technology capable to measure the copy number of tens of thousands of RNA species simultaneously within a cell [81]. The output of a scRNA-seq experiment can be intuitively understood as a count matrix where the rows identify individual cells, and the columns identify the genes in the cell's genome. While this data can be very useful to investigate the gene expression patterns associated to different types of cells, recently several models have been proposed to infer the regulation between genes [82,83]. Here, we briefly introduce the topic and focus on the emerging mathematical challenges. The inference of GRN from the scRNA-seq data introduces the challenge of inferring causal relationships from data that lacks temporal information. Generalizing the mathematical models of GRN developed in previous sections, the GRN inference problem can be generally set up as a nonlinear, high-dimensional system of the form

$$\dot{X} = f(X). \quad (41)$$

In eq. (24), $X$ represents the vector of mRNA counts of the $N$ genes in the cell, and $f(X)$ is a general nonlinear function that describes how genes mutually regulate each other. In most existing applications, the problem is further simplified by assuming linear interactions between genes and degradation

$$\dot{X} = AX - \gamma X, \quad (42)$$

where $A$ is a $N$x$N$ matrix encoding linear interactions between genes, with the implicit assumption that the positive and negative coefficients $A$ correspond to activation and inhibition, and $\gamma$ is a degradation rate vector. Interestingly, the existing bioinformatics methods for GRN inference can be broadly divided into two main classes based on whether steady state is assumed in eq. (42), with both approaches exhibiting strengths and pitfalls. First, steady state methods assume that $\dot{X} = 0$ [84,85]. Without further assumptions, however, this leads to the trivial, non-interacting solution $A = I\gamma$, where $I$ is the $N$x$N$ identity matrix. This problem is typically circumvented by eliminating self-interactions and setting the diagonal elements of $A$ to zero. With this strategy, the off-diagonal elements of $A$ can be determined, one row at a time, by solving the regression problem

$$\dot{x}_i = \sum_{j \neq i} a_{ij} x_j - \gamma_i x_i. \quad (43)$$

Bocci and collaborators showed that this method can detect the presence of GRN edges (i.e., causal connections between genes), but might lead to inaccurate sign prediction [28]. Alternatively, eq. (42) can be interpreted as an out-of-equilibrium problem [86–88]. This approach, however, requires an estimation of the first derivative $\dot{X}$ from scRNA-seq data that typically does not have temporal information. Thus, physical time is typically substituted with pseudotime. Pseudotime (pst) inference is a cell lineage reconstruction technique based on cell-cell gene expression similarity, where the pseudotime coordinate approximates real time and indicates the positions of cells along the lineage [89,90]. For example, in a differentiation process from multipotent toward differentiated cells, stem cells will be at the beginning of the lineage and thus will have smaller pseudotime coordinates, whereas the terminal cells will be toward the end of the lineage and thus will have large pseudotime values. After substituting physical time with pseudotime in the RHS, eq. (42) can be solved for $A$. While pseudotime is a powerful tool to reveal the lineages of biological systems, there might be pitfalls when using it as a proxy for physical time. For example, in a scenario of multistability cells at different points of a differentiation trajectory coexist at the same physical time while exhibiting different pseudotime values. Interested readers can find more in-depth description of GRN inference methods and their application to biological datasets in topical reviews [82,83].

### 3. Examples from various biological contexts

In this section, we examine examples where the tools to model multistable regulatory networks have been successfully applied and unraveled new biological functions. First, we consider the epithelial-mesenchymal transition (or EMT in short), a trans-differentiation process that regulates cell motility in physiological processes and diseases. In our description of EMT, we further review mathematical approaches to describe the regulation of non-coding RNAs which play an essential part in the control of EMT. Second, we review theoretical and computational models of stem cell differentiation, along with theoretical approaches that rephrase multistable gene regulatory networks as a many-body problem. Finally, we examine models of cell-cell communications through the Notch signaling pathway. These models introduce spatial patterning and multicellular communication that arise when cells can exchange information with their neighbors.

## 3.1 Intermediate states in the epithelial-mesenchymal transition

### 3.1.1. Epithelial-mesenchymal transition in development and cancer

The epithelial-mesenchymal transition (EMT) is a complex biochemical and biophysical process where epithelial cells, typically characterized by strong adhesion to neighboring cells and apicobasal polarity, loosen their adhesion and gain motility [91]. EMT and its reverse, MET, play a fundamental role in several developmental and physiological processes, such as organogenesis and wound healing, when cells need to transiently acquire motility to travel and rearrange their spatial organization. Moreover, EMT is implicated in several aspects of cancer progression, including metastasis and resistance to therapies [92,93]. A particularly interesting topic in EMT is the existence of intermediate cellular states separating the epithelial and mesenchymal phenotypes. These hybrid epithelial/mesenchymal cells (called E/M for short in the rest of the section) retain both cell-cell adhesion typical of epithelial cells and migration potential typical of mesenchymal cells. For this reason, these hybrid E/M states play an important role in collective cell migration in both physiological and pathological processes [94–96]. In recent years, theoretical modeling helped targeting several elusive questions about hybrid epithelial/mesenchymal states: is the EMT spectrum discretized in a set of intermediate states? Are these states truly stable or just metastable intermediates? Are they redundant or do they serve different biological functions? In this section we explore mathematical models that applied a wide array of methods to seek answers to these poignant questions.

### 3.1.2. Multistability predicted by continuous models of a core EMT regulatory network

Experiments have revealed intricated and interconnected network of genes, transcriptional factors and protein that regulate the epigenetic and biophysical transformations associated with EMT. Nonetheless, mathematical modeling has shown that some of the main aspects of EMT including the existence of intermediate cell states can be captured by a simple model of a core gene regulatory network including two micro-RNAs (miR-34 and miR-200) and two transcription factors (Zeb and Snail). In this core circuit, miR-34 and miR-200 inhibit, and are in turn inhibited, by Zeb and Snail, respectively. Moreover, Zeb regulates its own activity, while Snail activates Zeb and self-inhibits (Figure 8A). Zeb and Snail are typically activated when a cell undergoes EMT and are therefore associated with a mesenchymal phenotype. Conversely, miR-34 and miR-200 inhibit the action of Zeb and Snail and are therefore associated with an epithelial phenotype. A continuous framework proposed by Zhang and collaborators [97] models all these coupled interactions with Hill functions. Under this approximation, both the miR-34/Snail and miR-200/Zeb motifs behave as bistable toggle switches. Therefore, EMT is achieved by first flipping the miR-34/Snail switch from (high miR-34, low Snail) to (low miR-34, high Snail), and subsequently flipping the miR-200/Zeb switch from (high miR-200, low Zeb) to (low miR-200, high Zeb). The intermediate state with (low miR-34, high miR-200, high Snail, low Zeb) does not exhibit clear epithelial or mesenchymal signatures and is therefore interpreted as a hybrid E/M transition state (Figure 8B).

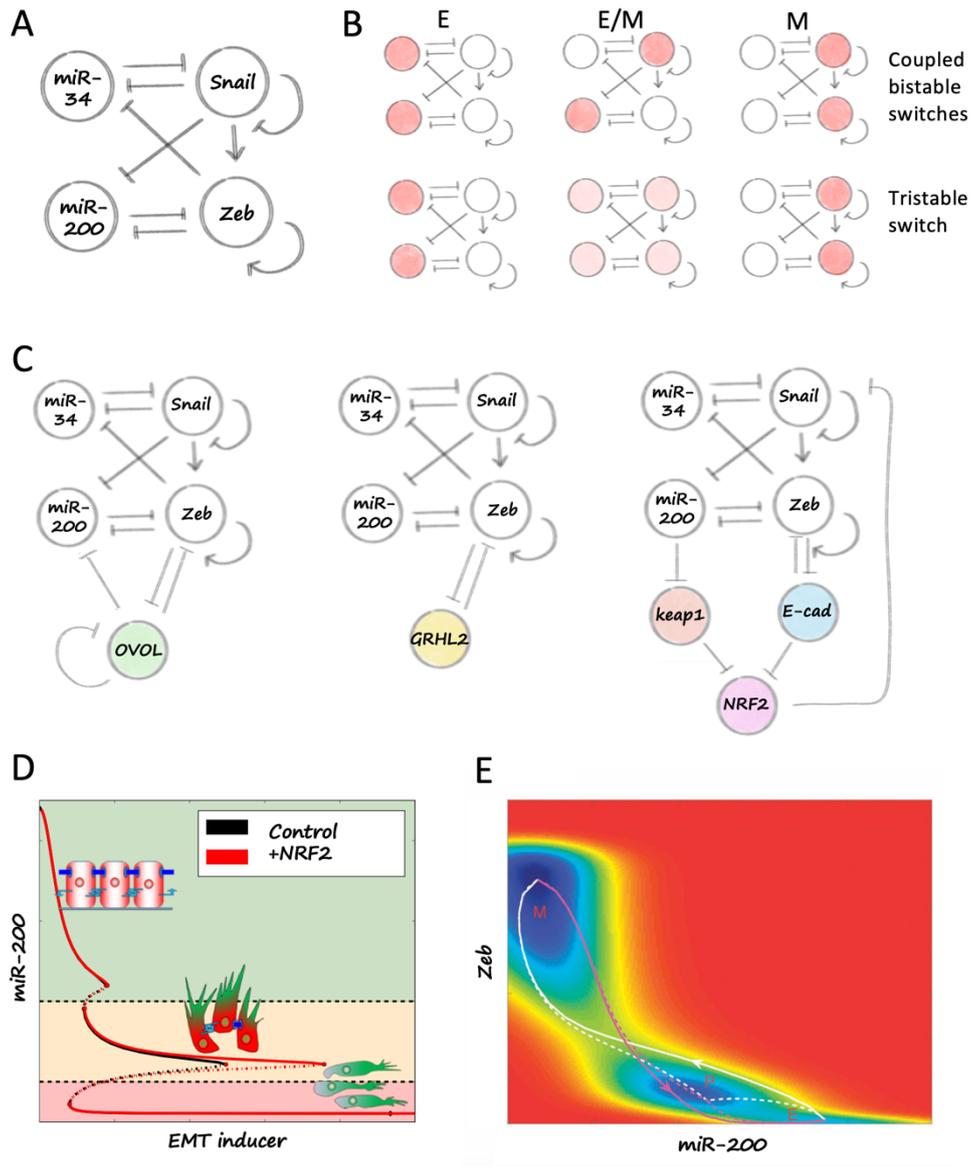

*Figure 8. **Continuous models of EMT.** (A) A core EMT gene regulatory network comprising the epithelial microRNAs miR-34, miR-200 and the mesenchymal transcription factors Zeb, Snail. (B) In the model by Zhang and collaborators [97], the transition state with (low miR-34, high miR-200, high Snail, low Zeb) is identified as the hybrid E/M. Conversely, in the model by Lu and collaborators [76], the hybrid E/M state has intermediate expression of all nodes. (C) Example of 'phenotypic stability factor' motifs that increase the stability of the hybrid E/M state by coupling to the core EMT circuit. (D) The stabilizing action of the PSF NRF2 is evaluated with a bifurcation diagram of miR-200 as a function of EMT inducer (adapted from Bocci and collaborators [98]); PSFs extend the hybrid E/M branch. (E) Pseudopotential landscape of the EMT circuit adapted from Li and collaborators [99]. E, P*

*and M indicate the locations of the epithelial, hybrid E/M (partial) and mesenchymal attractors. White and purple continuous lines indicate the minimum action paths (MAPs) of EMT and MET, while the dashed lines indicate the EMT and MET MAPs that pass through the hybrid E/M attractor.*

3.1.3. Micro-RNA based regulation of EMT

As we previously discussed, Hill functions can effectively model transcriptional activation/inhibition where a transcription factor binds to DNA – either alone, as a dimer, or as an oligomer, to enhance or repress the transcription of its target gene (see Insight box 2). Micro-RNAs such as miR-34 and miR-200, however, inhibit their target at a post-translational stage by binding to the transcribed mRNA and facilitating its degradation. This reaction, which is always inhibitory, is therefore better described with an explicit model of microRNA-mRNA binding, rather than Hill functions [100,101]. Targeted mRNA molecules have a variable number ($n$) of available binding sites for the micro-RNAs. The higher the number of micro-RNA molecules bound to the mRNA, the larger the degradation rate of the complex. Therefore, micro-RNAs inhibit their target by preventing translation of the mRNA into protein. Lu and collaborators [76] developed a model of EMT regulation with similar circuit topology but different functional form to describe the action of miR-34 and miR-200 [76,102]. The micro-RNA based chimeric circuit approach models the chemical binding between micro-RNA and target mRNA molecules and the possible translation and degradation rates as a function of number of occupied binding sites. This mathematical model is discussed in detail in the insight box 3.

Similar to the model of Zhang and collaborators [97], Lu and collaborators [76] predicted the existence of a hybrid E/M state. Different modeling assumptions on micro-RNA regulation, however, are reflected into distinct features of the hybrid E/M state. In the model of Lu and collaborators, the miR-200/Zeb circuit acts as a tristable switch with an additional stable fixed point with intermediate expression of both miR-200 and Zeb (Figure 8B). The miR-34/Snail circuit, conversely, is monostable, and was proposed to act as a buffer that filter noise and confers robustness to the miR-200/Zeb switch [76].

This models has been further extended to investigate the coupling with various 'phenotypic stability factors' that modify the topology of the core EMT circuit and increase the stability of the hybrid E/M phenotype [98,103–105]. This approach shifts the focus from the underlying biology characterizing a specific EMT regulator, and successfully predict its qualitative impact on EMT based on how it is connected to the core EMT gene regulatory network (Figure 8C-D).

Interestingly, a study by Nordick and collaborators[106] recently demonstrated that the cooperative RNA degradation by microRNAs alone can generate intermediate EMT states. The combination of such cooperative RNA degradation with transcriptional regulation can give rise to a broader spectrum of up to 7 states, which could better represent the broader spectrum of EMT states/EMT continuum recently observed by high-throughput transcriptomics measurements[107].

### 3.1.4. Landscape and path integral analysis highlights EMT transition routes

To gain further insights about the stability of the various EMT states and the transitions between them, Li and collaborators [99] studied the pseudopotential landscape of the EMT circuit of Lu and collaborators by explicitly solving the Fokker-Planck equation associated with the circuit. Specifically, the authors focused on the miR-200/Zeb switch that is responsible for the tristability of the circuit. This way, the pseudopotential can be easily visualized on a two-dimensional (miR-200, Zeb) space. The pseudopotential is obtained from the steady state probability $U = -\log\left(p_{SS}(miR-200, Zeb)\right)$ and is shown in Figure 8E. As expected, the landscape features three attractors corresponding to epithelial, hybrid E/M, and mesenchymal phenotypes. The authors further identified the minimum action paths (MAPs) connecting the three states. As discussed previously, the transitions are determined by the pseudopotential as well as the curl flux arising from broken detailed balance. Due to the curl flux, the MAP connecting the E and M attractors does not pass through the saddle points of the landscape, thus deviating from the behavior of 'classical', gradient-driven physical systems. There are several interesting biological implications stemming from this observation. First, cells undergoing E-to-M transition do not necessarily pass through the hybrid E/M phenotype, because the action of the E-E/M-M path is higher than the action of the direct E-M path. Moreover, the paths for the E-M transition and its reverse, M-E, are different, hence making them irreversible (Figure 8E). This suggests that cells undergoing EMT and MET might exhibit different molecular signatures. Finally, the action required for the E-E/M-M transition is higher than the action of the irreversible E-M transition; the action required for the first step (E-E/M transition), however, is lower than that of the E-M transition. Therefore, the E-E/M transition path could be chosen when the cell is not exposed to enough stimulus to undergo a complete E-M transition.

3.1.5. Boolean networks identify biological regulators and intermediate states in EMT

While small circuits of a core regulatory network enable a detailed analysis of EMT, several molecular players and signaling pathways participate in the regulation of EMT. Steinway and collaborators [108] reconstructed a large EMT Boolean network with 70 nodes and 135 connections by integrating experimental observations on known EMT factors and regulators from human hepatocellular carcinoma (HHC). To simulate the dynamics of EMT driven by TGF-$\beta$ – a well-known EMT inducer – the authors initialized the network in an epithelial state and updated the model following a stochastic asynchronous scheme. To decrease the complexity of the network, they also applied reduction methods to eliminate 'redundant' nodes and connections. Interestingly, they identified a reduced network with 19 nodes and 70 connections with a very similar dynamical behavior (Fig. 9A). This reduced network enabled the identification of transition trajectories from the initial epithelial state to the final mesenchymal state, thus offering interesting information on the temporal ordering in the activation and deactivation of relevant genes. Furthermore, the authors explored the effect of knockout and/or constitutive activation of nodes or combinations of nodes by artificially enforcing an 'ON' or 'OFF' state for certain nodes in the circuit [109]. Strikingly, some of these perturbations prevent EMT but do not reverse the transition to the original epithelial attractor; rather, alternative attractors with both epithelial and mesenchymal active genes become stable, which are identified as candidates for hybrid epithelial/mesenchymal states.

Font-clos and collaborators [110] described the same EMT network with a pseudo Hamiltonian based on the formalism of spin systems $H = -\sum_{i,j} J_{ij} s_i s_j$. Compared to the Hamiltonian typically used to describe spin systems, here the $s_i = 0, 1$ describe an inactive or active gene, respectively, and the $J_{ij}$ describe inhibition ($J_{ij} = -1$), activation ($J_{ij} = 1$) or lack of interaction ($J_{ij} = 0$) from gene $i$ to gene $j$. Thus, differently from standard spin system Hamiltonians, interactions are not symmetric (i.e., $J_{ij} \neq J_{ji}$) and therefore it is not guaranteed that the fixed points of the system correspond to the minima of $H$. The authors sampled the landscape of the network with a large number of stochastic simulations using the same update scheme originally used by Steinway and collaborators [108,109]. This analysis revealed two deep minima associated with the epithelial and mesenchymal states separated by a complex landscape with several local minima that are interpreted as intermediate, less stable states. On a two-dimensional PCA map, the distribution of steady states shows two dense regions corresponding to epithelial (E-cadherin node is ON) and mesenchymal (E-cadherin node is OFF), separated by a sparser region where E-cadherin can be ON or OFF (Fig. 9B-C). This distribution is reminiscent of frequencies and relative stabilities of epithelial, mesenchymal and hybrid epithelial/mesenchymal states as identified in multiple Boolean models of EMT regulatory networks [111].

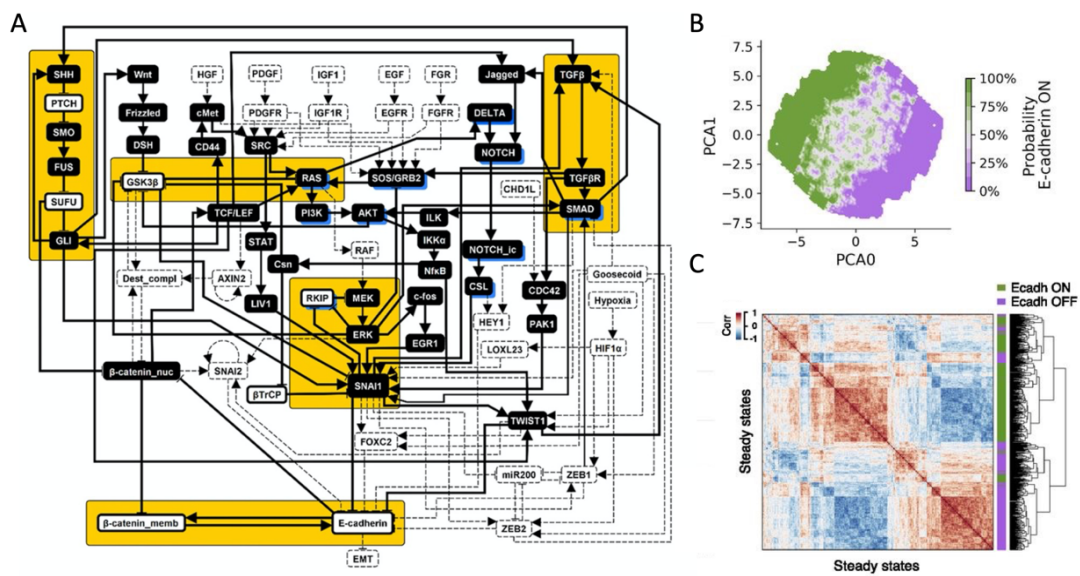

*Figure 9. **Boolean models of EMT.** (A) A large EMT network used by Steinway and collaborators [109]. Yellow areas highlight the motifs that are found to stabilize the epithelial state. White-colored nodes are active in the epithelial state while black-colored nodes are inactive. Nodes with blue background are predicted as sensible knockout targets*

*that suppress EMT. (B) Density of steady states on the PCA0-PCA1 plane in the model of Font-Clos and collaborators [110]. Green and purple color shading indicate high or low probability that E-cadherin is on in the steady state. (C) Clustering and correlation of 500 steady state solutions in the model of Font-Clos and collaborators.*

---

3.1.6 Insight box 3: Describe micro-RNA mediated inhibition with mass action and chimeric circuits

Micro-RNAs (miRs) inhibit the expression of a protein by binding to the target mRNA molecules and degrade them before translation. Several models that describe miR-mediated inhibition employ the principles of mass action [100,101]. Suppose that mRNA of a given protein ($m$) is transcribed with rate $k_m$ and the corresponding molecule ($p$) is translated with rate $k_p$. Moreover, we consider a miR species ($\mu$) that is produced and degraded with rate constants $k_\mu$, $\gamma_\mu$, respectively. $\mu$ binds and unbinds to $m$ with rate constants $k_B$ and $k_U$, respectively, and the mRNA-miR complex $[m\mu]$ degrades with rate constant $\gamma_{m\mu}$. The coupled dynamics of mRNA, miR, mRNA-miR complex and protein ($m, \mu, [m\mu], p$) is described by the following system of ODEs

$$\frac{dm}{dt} = k_m - k_B m\,\mu + k_U [m\mu] - \gamma_m m, \quad (44a)$$

$$\frac{d\mu}{dt} = k_\mu - \gamma_\mu \mu, \quad (44b)$$

$$\frac{d[m\mu]}{dt} = k_B m\,\mu - k_U [m\mu] - \gamma_{m\mu} [m\mu], \quad (44c)$$

$$\frac{dp}{dt} = k_p m - \gamma_p p. \quad (44d)$$

Various approximations can be made from this basic framework, for instance by assuming fast mRNA-miR binding and writing the loss term of $m$ as a function that only depends on $m$ and $\mu$. Moreover, as discussed in the insight box 1, fast mRNA dynamics can be another valid approach. More details on this are provided in other reviews [100,101].

The general scheme of eqs. (44), however, assumes a 1-to-1 stoichiometry in the mRNA-miR binding. Often, mRNA molecules have multiple binding sites, whose occupancy modulates translation and degradation rates (Figure 10). In the following, we discuss a generalization of eqs. (44) developed by Lu and collaborators that accounts for multiple miR binding sites that has been successfully applied to the gene circuit that regulates the epithelial-mesenchymal transition [76,102].

Similar to eqs. (44), let us consider a system with an mRNA species ($m$) and an inhibiting miR ($\mu$). Moreover, let us assume that each mRNA molecule has $n$ binding sites for miR molecules. miR molecules bind and unbind to the mRNA with rate constants $r_{\mu+}$ and $r_{\mu-}$, respectively. In the model, binding at different sites are treated as independent events due to the small size of the miR compared to the mRNA molecule. Assuming fast equilibration of mRNA-miR binding, the fraction of mRNA molecules with $i$ occupied binding sites satisfies

$$r_{\mu+} \, \mu \, [m_i] = r_{\mu-} \, [m_{i+1}]. \quad (45)$$

Iteratively, $[m_i]$ can be expressed as a function of $[m_0]$, i.e. the fraction of mRNAs without any occupied binding site

$$[m_i] = \left(\frac{\mu}{\mu_0}\right)^i [m_0], \quad (46)$$

where $\mu_0 = r_{\mu-}/r_{\mu+}$. The set of concentrations $[m_i]$, $i = 0, 1, \ldots n$, must also satisfy:

$$\sum_{i=0}^{n} C_n^{\,i} \, [m_i] = m, \quad (47)$$

where $C_n^{\,i} = n!/i! \, (n-1)!$ is the number of arrangements for $i$ micro-RNA molecules on $n$ binding sites. Plugging eq. (46) into eq. (47) yields

$$m = [m_0] \sum_{i=0}^{n} C_n^{\,i} \left(\frac{\mu}{\mu_0}\right)^i = [m_0] \left(1 + \frac{\mu}{\mu_0}\right)^n. \quad (48)$$

Thus, using eq. (46) and (48), the fraction of mRNA molecules with $i$ miR bound ($[m_i]$) can be expressed as

$$[m_i] = m \frac{\left(\frac{\mu}{\mu_0}\right)^i}{\left(1 + \frac{\mu}{\mu_0}\right)^n}. \quad (49)$$

Finally, we introduce sets of rate constants $[l_i]$, $[\gamma_{mi}]$ and $[\gamma_{\mu i}]$ that represent the translation rate of a mRNA molecule with $i$ miR bound, the degradation rate of a mRNA molecule with $i$ miR bound, and the degradation rate of miR molecules associated with $i$-bound mRNAs. Therefore, it is possible to write total translation and degradation terms

$$\frac{dm}{dt} = k_m - m \sum_{i=0}^{n} \gamma_{mi} \, C_n^{\,i} \, M_n^{\,i}(\mu) - \gamma_m m, \quad (50a)$$

$$\frac{d\mu}{dt} = k_\mu - m \sum_{i=0}^{n} \gamma_{\mu i} \, C_n^{\,i} \, M_n^{\,i}(\mu) - \gamma_\mu \mu, \quad (50b)$$

$$\frac{dp}{dt} = k_p m \sum_{i=0}^{n} l_i \, C_n^{\,i} \, M_n^{\,i}(\mu) - \gamma_p p, \quad (50c)$$

where $M_n^{\,i}(\mu) = \left(\frac{\mu}{\mu_0}\right)^i \Big/ \left(1 + \frac{\mu}{\mu_0}\right)^n$.

Compared to modeling inhibition with Hill functions or simpler mass-action dynamics, the chimeric circuit approach requires knowledge of more parameters which might or might not be readily accessible from experiments. In the case of the EMT circuit proposed by Lu and collaborators, the expression levels of translated protein with various number of occupied mRNA binding sites were used to calibrate the model's parameters. Further information about this procedure can be found in the original publication [76].

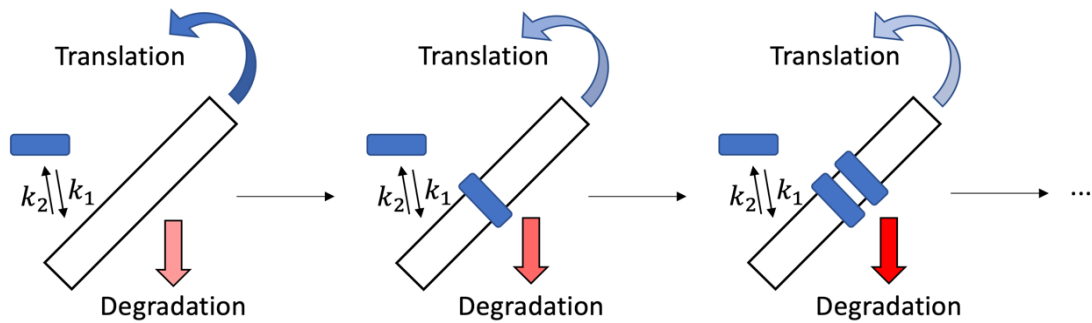

*Figure 10.* **Schematic representation of microRNA-based regulation in the chimeric circuit model.** *A mRNA molecule (large white-filled rectangle) can bind to multiple microRNA molecules (small blue rectangles). mRNA-microRNA binding does not depend on the number of already occupied binding sites (same binding and unbinding rate constants in the different pictures). The more microRNA molecules bound, the higher the complex degradation rate*

*(indicated by brighter red arrow from leftmost to rightmost panel) and the lower the mRNA translation rate (indicated by lighter blue arrow from leftmost to rightmost panel).*

---------------------------------------------------------------------------------------------------------------------

**3.2 Stem cell differentiation**

It has been a continuing research interest to understand how stem cells accurately specify cell fates during differentiation, referred to as stemness regulation. Mathematical modeling approaches have been applied extensively to elucidate the gene regulatory mechanisms underlying stem cell differentiation.

3.2.1 A toggle switch topology regulates the differentiation of stem cells

Mathematical models focusing on the acquisition of stable states representing cell phenotypes are often derived using chemical rate equations. Similar to the approaches discussed in EMT that focus on small circuits, the modeling of stem cell differentiation are often applied to the proposed "core" stemness regulatory networks that contain master stemness regulators.

Huang and co-workers developed a continuous model focusing on the binary decision-making during the lineage commitment of erythroid and myelomonocytic fates[112]. The model is derived to represent the temporal dynamics of a core circuit containing the two lineage specifying transcription factors (TFs) GATA1 and PU.1. GATA1 and PU.1 exhibit mutual inhibition and auto-stimulation, which are modeled by Hill functions (Fig. 11A). Modeling analysis shows that the circuit can generate stable attractors corresponding to erythroid (GATA1 high, PU.1 low) and myelomonocytic (GATA1 low, PU.1 high) phenotypes, and a metastable state representing the "multilineage priming" stage, characterized by coexpression of both GATA1 and PU.1. This study describes a binary cell fate decision-making for a bipotent progenitor cell.

Another kinetic model of stemness regulation has been developed by Jolly and co-workers focusing on the LIN28/let-7 circuit[113]. LIN28 (an RNA-binding protein) and let-7 (a microRNA) are mutually inhibitory and self-excitatory (Fig. 11B). Two ordinary differential equations have been derived to represent the birth and death process of LIN28 and let-7. The shifted Hill functions are used to represent the let-7/LIN28 mediated inhibition and self-activation. The modeling analysis suggests that the LIN28/let-7 circuit can operate as a three-way switch enabling three stable states characterized by (high LIN28, low let-7), (high let-7, low LIN28) and (intermediate LIN28, intermediate let-7). As the experimental studies suggest that intermediate levels of OCT4 account for pluripotency and OCT4 is a downstream target of LIN28, this model proposes the stable state characterized by (intermediate LIN28, intermediate let-7) associates with pluripotency. From a theoretical perspective, this model highlights the effect of additional regulatory interactions on the "standard" toggle switch topology. While mutual inhibition typically leads to bistability, the additional self-activation of both LIN28 and let-7 introduces a third state with intermediate expression of both factors.

To elucidate the transcriptional dynamics of the master stemness regulatory TFs - OCT4, SOX2 and Nanog during stem cell differentiation, Chickarmane and co-workers developed a kinetic model to study the stemness circuit containing three master stemness regulatory TFs - OCT4, SOX2 and Nanog[114] (Fig. 11C). These three TFs play a critical role in positively regulating the stemness genes and negatively regulating the differentiation genes. The modeling analysis in this study shows that the OCT4/SOX2/Nanog circuit can function as a binary switch, and enable two stable states that are characterized by (high OCT4-SOX2, high Nanog) and (low OCT4-SOX2, low Nanog), respectively. Here OCT4-SOX2 is a heterodimer formed by OCT4 and SOX2 and can regulate OCT4, SOX2 and Nanog individually. The stable state characterized by (high OCT4-SOX2, high Nanog) exhibits high expression of stemness genes and low expression of differentiation genes thus corresponding to stem cell phenotype. Conversely, the stable state characterized by (low OCT4-SOX2, low Nanog) corresponds to cells undergoing differentiation. Through bifurcation analysis, the study shows how stem cells - (high OCT4-SOX2, high Nanog) can maintain their stemness state upon removal of the stimulus signals that upregulate OCT4 and SOX2. Later Chickarmane and Peterson [115] extended their model by including additional stemness TFs (Cdx2, Gcnf, Gata-6) to elucidate the Trophectoderm and Endoderm lineage commitment. By performing perturbation analysis on the circuit, the study suggests strategies to reprogram cells back to stemness state, such as activation of Nanog and suppression of Gata6.

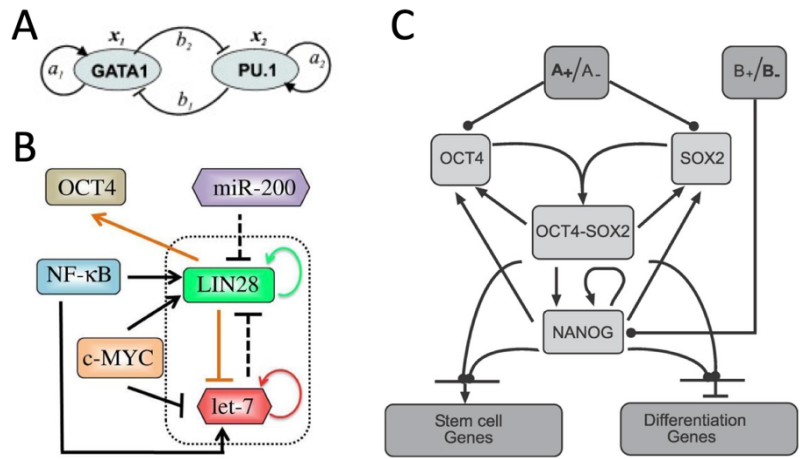

*Figure 11. **Core regulatory circuits to model stem cell differentiation.** (A) Huang and coworkers* [112] *modeled stem cell differentiation via mutual inhibition between GATA1 and PU.1. (B) The stemness circuit constructed by Jolly and collaborators* [113]. *to model the acquisition of cancer stem cell traits in cancer. (C) The circuit developed by Chickarmane and co-workers* [114] *focuses in the stemness regulators OCT4, SOX2 and Nanog.*

3.2.2 Role of stochastic fluctuations in stem cell differentiation

While the deterministic models discussed in the previous section successfully capture the stem cells decision-making dynamics, it remains unclear how this differentiation is regulated in the fluctuating cellular environment. To capture the stochastic effects, Kalmar and co-workers developed a continuous model based on the Oct4/Nanog circuit[116]. The dynamics of the Oct4/Nanog circuit are represented by two coupled differential equations. The transcription noise (represented by Gaussian white noise) is incorporated into the differential equation representing the temporal dynamics of Nanog. This study sheds light on how cells can vary their Nanog level without losing their pluripotency. The study suggests an interesting view of pluripotency. The pluripotency, instead of being viewed as a stable state characterized by fixed amount of gene expression levels, can be viewed as a heterogenous population driven by transcriptional noise such as the noise during Nanog transcription.

To continue studying the mechanisms underlying the heterogenous levels of Nanog in mouse embryonic stem cells, Glauche and co-works developed an ODE model to simulate the Oct2/Sox2/Nanog circuit[117]. They showed that the experimental characteristics of Nanog variation can be recapitulated by the model via either adding a Gaussian white noise term in the ODE representing Nanog (noise-induced transition between Nanog low and Nanog high states) or including a transcriptional repressor of Nanog (oscillation). This study also shows how Nanog low cells can be more prone to differentiation relative to Nanog high cells. Another modeling study to elucidate the variation of Nanog levels in embryonic cells has been performed by Herberg and co-workers by integrating a negative feedback loop FGF4/Erk signaling with the Oct4/Sox2/Nanog/Rex1 circuit. The effect of the transcriptional noise, defined by the zero-mean Gaussian process, has been incorporated into the ODEs representing the temporal dynamics of Oct4-Sox2, Nanog and Rex1. The model shows a bimodal distribution of Nanog levels and blocking the Erk signaling pathway can lead to a merged Nanoga distribution[118].

### 3.2.3 Chemical master equation models characterize the role of Nanog during differentiation

The heterogeneous expression of Nanog in pluripotent cells has also been studied by applying master equations to describe the stochastic dynamics of a stemness network containing eight stemness TFs (Oct4, Sox2, Nanog, GATA6, Gcnf, Cdx2, Pbx1, Klf4) and one heterodimer (Oct4-Sox2) formed by Oct4 and Sox2[119]. The master equations for the stemness network were derived by considering both the concentration of proteins and the occupancy of the DNA sites. This study shows how the stemness network can enable up to five steady states, corresponding to two stem cell phenotypes expressing (high Oct4, high Sox2 and high Klf4), one primitive endoderm phenotype, one trophectoderm phenotype and one differentiated cell type. The stem cell steady states are differentiated by distinct levels of Nanog, which is consistent with its experimental characteristics. Through analyzing the most probable transition paths among the steady states, the study suggests a sequential transition path that *Nanog high* stem cell steady state first decreases the Nanog levels to enter a novel, *Nanog low* stem cell steady state, and then transition to the primitive endoderm steady state. The modeling results support the critical role of Nanog in safeguarding stemness against differentiation.

Furthermore, to overcome the computational limitations of simulating large networks, Sasai and Wolynes described the stochastic gene expression and feedback interactions between transcription factors using the many-body approach typically employed in quantum statistical mechanics [20]. In the Insight box 4, we introduce this approach and show how the chemical master equation of an autoregulating transcription factor can be effectively mapped with this formalism. By extending this approach to an 8-gene circuit of embryonic stem cell development, they characterized the multistable landscape and the most probable transitions between attractors [119].

3.2.4 Landscape theory identifies the trajectories of differentiation and reprogramming
To quantify the kinetic flow during embryonic stem cell differentiation, epigenetic landscape approach has been applied. Li and Wang developed a global potential landscape for a stem cell development network containing fifty-two genes[120]. There are two basins of attractions identified on the landscape corresponding to a stem cell state and a differentiated state. The model shows the development process can be viewed as a transition from the stem cell attractor to the differentiated attractor. Interestingly, the reversed "reprograming" trajectory where differentiated cells acquire pluripotent traits can be different in the 52-gene expression space. Later, the authors applied the landscape and flux path approach to quantify the landscape for a reduced cancer stem cell (CSC) circuit that only contains six key genes (two microRNAs - miR200 and miR-145, four proteins - ZEB, P53, MDM2 and OCT4), to identify the basins of attractions and transition paths in between[121]. This study shows the CSC circuit enables four basins of attractions corresponding to a normal state, a normal stem cell state, a cancer state, a cancer stem cell state. Specifically, the study suggests that P53 activation promotes the transition from the cancer stem cell attractor to the normal cell attractor, which is consistent with the tumor suppressor role of P53.

To understand the large cell-to-cell variation of Nanog and other stemness TFs in embryonic stem cells, Sasai and co-workers developed an epigenetic landscape approach for a stemness regulatory network contains six genes - three stemness genes (Sox2, Oct4, Nanog), and three lineage-specific genes (Gata6, Cdx2, and Gcnf) [122]. One interesting prediction from this study is the distribution of time scales of the regulatory processes (binding/unbinding of TFs, histone modification etc.) is critical to determine the dynamic behaviors of the network. Specifically, the slow transcriptional switching of Nanog leads to large fluctuations among multiple transient states.

3.2.5 Parameter randomization on the stemness circuit

To identify the robust gene states during stem cell differentiation, parametric randomization methods have been applied to the stemness network. For example, RACIPE has been applied to a proposed core stemness regulatory network containing eight master regulatory TFs (Oct4, Sox2, Cdx2, Gata6, Gcnf, Pbx1, Klf4 and Nanog) [123]. The stemness regulatory network contains mostly transcriptional regulation with protein-protein interaction, i.e., the binding/unbinding interactions between SoX2, Oct4 and the Oct4-Sox2 complex. Using exclusively the topological information of the network as input, RACIPE generated an ensemble of mathematical models with various numbers of steady-state solutions that can be explored with statistical analysis. Hierarchical clustering results of the RACIPE solutions exhibit distinct gene expression patterns that can be associated with different development stages and match the single-cell mRNA expression data of mouse embryonic stem cells (ESCs). One intriguing result of this study is the multi-stable behaviors of the stemness GRN and the recapitulation of the experimentally measured gene expression patterns cannot be achieved by networks with randomized topologies, even though the randomized networks can have similar amount of mutual inhibitory and self-excitatory feedback loops relative to the stemness circuit. The results suggest that the stemness GRN may indeed contain some topological properties that are beyond the expectation based on the counting of the simple motifs. By analyzing the difference of the parameters that lead to different states, the key parameters whose variation can induce the transitions among different states are identified. The physiological representation of those parameters can be the stimuli that trigger the transition. Through systematically perturbing each gene and each regulatory link in the stemness GRN, RACIPE elucidates a hierarchical decision-making structure of the stemness GRN, with the OCT4/CDX2 motif functioning as the first decision-making module followed by the GATA6/Nanog module. By simulating the effect of external signals that perturb the TFs in the stemness GRN, RACIPE analysis results suggest that the presence of external signals often restrains the gene states that can be accessed instead of creating new states. The results suggest an alternative explanation of the Waddington landscape that the stem cell population, instead of being viewed as a fixed stable "stemness" state, can be regarded as a mixture of heterogeneous cell phenotypes. Along the differentiation, upon external signals, the cell population heterogeneity decreases partially because cells lose access to some stable states. In other words, cells are committed to specific phenotypes, therefore, differentiated.

3.2.6 Combing Landscape theory with random parametric perturbation

To calculate the landscape of stem cell networks by considering the large variation of parameters, Li developed an approach that combines landscape theory with random parameter perturbation, referred to as LRPP[124]. The LRPP approach contains the following steps: the time evolution of each gene in the stemness network is represented by ODEs; the modeling parameters are sampled from a range with either uniform or Gaussian distribution; multiple initial conditions are chosen to identify all possible steady states; the landscape is calculated using the self-consistent mean field approximation; repeat the aforementioned procedure for multiple times and Sum over the landscape of different sets of parameters to get the landscape for the ensemble models. By applying the LRPP approach to a stemness GRN that contain four stem cell marker genes OCT4, SOX2, NANOG, and KLF4 and three differentiation marker genes GCNF, CDX2, and GATA6, the author identifies three main stable basins of attractions corresponding to an embryonic stem cell state (high Nanog, low Gata6), a differentiate cell state (low Nanog, high Gata6) and an intermediate state (high Nanog, high Gata6). The intermediate state characterized by co-expression of stemness marker Nanog and differentiation marker Gata6 may account for the heterogeneity of these TFs observed during single-cell experiments, and may play a critical role in regulating stem cell plasticity.

--------------------------------------------------------------------------------------------------------------------

3.2.7 Insight box 4: spin-boson formalism for an autoregulatory gene

In this insight box, we discuss the application of spin-boson formalism to describe stochastic gene switches. Here, we only aim at giving the reader a flavor of this approach. Following

Sasai and Wolynes, we examine a single autoregulatory gene (i.e., a gene that regulates its own transcription). An extensive review on the uses of quantum field theory in diffusion-reaction systems is offered by Mattis and Glasser[125].

The autoregulatory gene is described by the protein copy number $n$ and a Boolean variable $s = 0,1$ describing whether the gene is bound or unbound.

We have previously developed the CME for a birth-death process (see insight box 1). In addition to production and degradation, produced molecules can bind to DNA and regulate their own transcription. Depending on whether molecules bind to DNA as monomers or oligomers, the protein-DNA binding rate depends on a power of protein level $n$ (see insight box 2). Since we are not interested in solving a specific case but rather layout the mapping procedure, we assume a generic binding rate $h(n)$. The DNA-protein unbinding rate is a constant $f$ independent from $n$. The unbound and bound promoter produces molecules at rates $g_0$ and $g_1$, respectively. Similar to the birth-death process, we are condensing mRNA transcription and translation into a single production rate. Finally, protein molecules degrade with rate constant $k$. A set of two probabilities $\big(P_1(n,t), P_0(n,t)\big)$ describes the probabilities to have $n$ molecules and an unbound/bound promoter at time $t$, respectively. Combining all reactions, CMEs for $P_0$ and $P_1$ assume the form

$$\frac{dP_0(n)}{dt} = g_0 P(n-1,0) - g_0 P(n,0) + k(n+1)P(n+1,0) - knP(n,0)$$
$$+ fP(n,1) - h(n)P(n,0), \quad (51a)$$

$$\frac{dP_1(n)}{dt} = g_1 P(n-1,1) - g_0 P(n,1) + k(n+1)P(n+1,1) - knP(n,1)$$
$$- fP(n,1) + h(n)P(n,0). \quad (51b)$$

To map this CME onto a many-body problem, the first step is to rearrange eqs. (51) as a single equation for the probability vector $P(n,t) = \big(P_1(n,t), P_0(n,t)\big)$

$$\frac{\partial}{\partial t}P(n,t) = \begin{pmatrix} g_1 & 0 \\ 0 & g_0 \end{pmatrix}[P(n-1,t) - P(n,t)] + k[(n+1)P(n+1,t) - nP(n,t)]$$
$$+ \begin{pmatrix} -h(n) & f \\ h(n) & -f \end{pmatrix} P(n,t). \quad (52)$$

Further, we define a state vector $\psi$:

$$\psi = \sum_{n=0}^{+\infty} P(n,t)\,|n\rangle. \quad (53)$$

To make the connection to the quantum many-body problem more evident, we introduce ladder operators[125]

$$a^*|n\rangle = |n+1\rangle, \quad (54a)$$
$$a|n\rangle = n|n-1\rangle. \quad (54b)$$

Notably, the ladder operators still satisfy the commutation relation $[a, a^*] = 1$, but their definition slightly differ from the standard harmonic oscillator operators ($a^*|n\rangle = \sqrt{n+1}|n\rangle$, $a|n\rangle = \sqrt{n}|n-1\rangle$). This unconventional definition arises because the coefficients in the expansion of eq. (53) are probabilities, whereas typically probabilities correspond to the squared coefficients in 'standard' expansions of the wave function.

With these definitions in hand, the CME can be rewritten in the form

$$\frac{\partial \psi}{\partial t} = \Omega \psi, \quad (55)$$

where all the information about chemical reactions is encoded by the operator $\Omega$:

$$\Omega = (\bar{g} + \delta g\, \sigma_Z)(a^+ - 1) + k(a - a^+ a) + \mu^+(-1 + \sigma_X) + \mu^-(-i\sigma_Y - \sigma_Z). \quad (56)$$

With the following definitions

$$\bar{g} = \frac{(g_1 + g_0)}{2}, \quad (57a)$$
$$\overline{\delta g} = \frac{(g_1 - g_0)}{2}, \quad (57b)$$
$$\mu^+ = \frac{(h(aa^+) + f)}{2}, \quad (57c)$$
$$\mu^- = \frac{(h(aa^+) - f)}{2}, \quad (57d)$$

and the introduction of the matrices:

$$\sigma_X = \begin{pmatrix} 0 & 1 \\ 1 & 0 \end{pmatrix}, \quad (58a)$$
$$i\sigma_Y = \begin{pmatrix} 0 & 1 \\ -1 & 0 \end{pmatrix}, \quad (58b)$$

$$\sigma_Z = \begin{pmatrix} 1 & 0 \\ 0 & -1 \end{pmatrix}, \quad (58c)$$

Sasai and Wolynes further apply this formalism to a more complex 8-gene circuit of stemness transcription factors to predict stable attractors and transitions corresponding to differentiation. More details about this work can be found in reference [20].

---

**3.3 From multistability to spatial patterning: cell communication through Notch signaling**

So far, we have considered intracellular signaling networks that regulate cell-fate dynamics. In all these cases, the decision on cell fate was cell-autonomous, i.e., it did not depend on the signaling state of other cells in the surrounding environment. A cell population where cells can assume one of multiple states will exhibit heterogeneity according to the states' relative stability but without any spatial organization. If cells exchange information with their neighbors, however, the cell-fate decision processes become correlated, thus giving rise to spatial organization. In this section, we consider the example of Notch signaling, one of the most well-conserved and studied signaling pathways that regulate both physiological and pathological processes [126,127]. Notch signaling operates via binding of ligands and receptors belonging to neighboring cells, thus serving as a nearest neighbor communication mechanism that couples cell-fate decisions in a spatially-dependent manner. Signaling through different classes of Notch ligands can lead to either converging or diverging cell states between neighbors, thus raising interesting parallels with the behavior of spin systems such as ferromagnets and antiferromagnets.

3.3.1 Notch signaling relays single cell multistability to spatial patterning

Notch signaling is initiated when the extracellular domain of the Notch receptor binds to the transmembrane domain of a ligand at the surface of a neighboring cell. Mammalian species typically exhibit one class of receptors (Notch) and two classes of ligands (Delta and Jagged), which can be further divided into a variable number of subtypes with different molecular structures. Upon binding, pulling forces by endocytosis and sequential cleavage actions by assisting enzymes lead to the release of the Notch intracellular domain (NICD). The NICD is transported to the cell nucleus where it activates or inhibits the transcription of several target genes [126,128,129]. In particular, NICD transcriptionally activates Notch and Jagged while inhibiting Delta. Therefore, a cell with high expression of Delta ligands activates the Notch receptors in its neighbors, thus in turn implying the repression of Delta. Conversely, when a cell is exposed to low levels of Delta from its neighbors, the Notch receptor is not activated, thus allowing production of Delta. Hence, Notch-Delta signaling leads neighboring cells to divergent cell states: a (low Notch, high Delta) state typically referred to as Sender, and a (high Notch, low Delta) state typically referred to as Receiver. At the multicellular level, this patterning principle leads to alternation of Senders and Receivers, typically referred as "lateral inhibition", which plays a crucial role in the differentiation of cell states in several physiological processes including somitogenesis, angiogenesis and neurogenesis [130–132]. Conversely, a cell with high expression of Jagged activates Notch receptors in its neighbors, which in turn activates the production of both Notch and Jagged. Therefore, Notch-Jagged signaling promotes a convergent (high Notch, high Jagged) state among neighbors that is often referred to as hybrid Sender/Receiver. On a multicellular level, this patterning principle leads to a homogeneous population of hybrid Sender/Receiver cells, or "lateral induction", which is observed, for instance, in the spatial propagation of a pluripotent cell state during inner ear development [130,131,133]. In the following paragraphs, we will review models that investigate how intracellular signaling and ligand-receptor binding relay cell-fate decisions in individual cells and multicellular patterning, the competition between lateral inhibition and lateral induction, and the role of stochastic fluctuations in enforcing or disrupting ordered patterns.

3.3.2 Notch-Delta lateral inhibition: a two-cell toggle switch

The first mathematical model of Notch signaling proposed by Collier and collaborators [134] directly generalizes the single cell toggle switch and focuses on lateral inhibition driven by Notch-Delta signaling between neighboring cells (Figure 12A). It considers a two-dimensional lattice where the temporal dynamics of Notch and Delta in a cell ($p$) is described by the set of ODEs

$$\frac{d(N_p/N_0)}{dt} = \frac{(\overline{D_p}/D_0)^k}{a + (\overline{D_p}/D_0)^k} - \mu(N_p/N_0), \quad (59a)$$

$$\frac{d(D_p/D_0)}{dt} = \frac{1}{1 + b(N_p/N_0)^h} - \rho(D_p/D_0), \quad (59b)$$

where $\overline{D_p} = \frac{1}{c}\sum_{p'} D_{p'}$ is the average level of Delta ligand in the nearest neighbors; in this expression $c$ is the number of nearest neighbors and the summation goes over all nearest neighbor cells $p'$. $N_0$ and $D_0$ are typical levels of Notch and Delta used to scale the model while $\mu, \rho$ are dimensionless degradation rate constants. Compared to all the circuits reviewed so far, this model considers a multicellular lattice (typically two-dimensional) where each individual cell is described by the Notch-Delta circuit, and circuits of neighboring cells are connected through the nearest neighbor summations in the Hill functions of eq. (59a).

Therefore, if neighbors of cell $p$ are Senders with high Delta, cell $p$ represses the production of Delta while increasing the production of Notch, hence assuming a (high Notch, low Delta), Receiver state. Conversely, if neighbors of cell $p$ are Receivers with low Delta, Notch is weakly activated in cell $p$, thus maintaining a Sender state with (low Notch, high Delta) [132,134]. Therefore, if two neighboring cells start with very similar but not exactly equal levels of Notch and Delta at time $t = 0$, the mutual inhibition mechanism will amplify the small initial difference and ultimately differentiate between a Sender cell and a Receiver cell.

On an extended two-dimensional lattice, this model allows patterns where Sender cells are surrounded by Receiver cells. Specifically, on a square lattice, these interactions lead to a chessboard-like pattern with alternating Senders and Receivers [135]. On a hexagonal lattice that perhaps better represents the arrangement of cells in an epithelial tissue, however, a perfect alternation of Senders and Receivers cannot be achieved due to lattice frustration. Therefore, Sender cells are typically surrounded by six Receivers that are in contact with one another, leading to a 3-to-1 Receiver/Sender ratio (figure 12B). Indeed, Receiver cells can be in contact as they do not express Delta, and therefore do not actively regulate each other, while Sender-Sender contacts give rise to mutual inhibition until one cell is converted to the Receiver state [136]. In other words, a contact between Receivers simply results in lack of cell-cell signaling, whereas a contact between Senders leads to the mutual inhibition that ultimately breaks the symmetry and forces one cell to the Receiver state. This mechanism is not flawless and sometimes patterning mistakes can be observed; in the last section of the chapter, we will discuss in more details the nature of these "defects", their biological implications and the role of stochastic fluctuations in modulating these patterns.

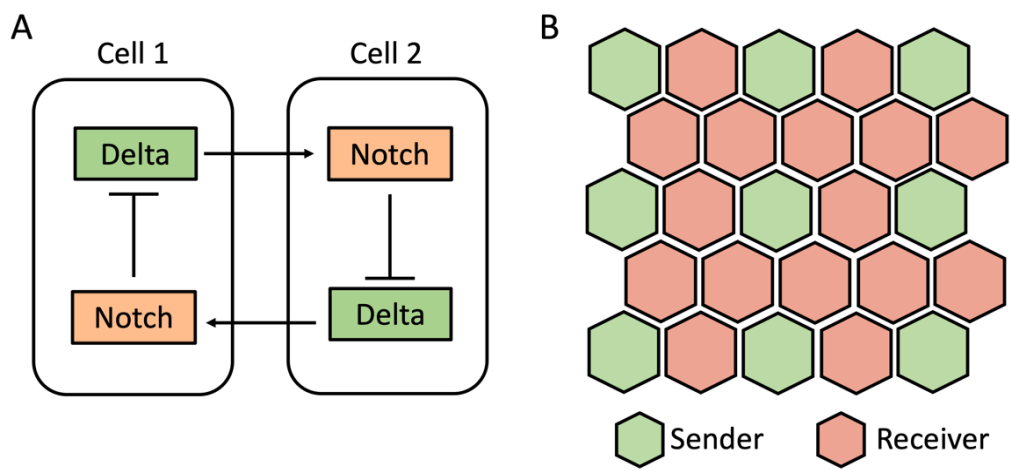

*Figure 12.* **Lateral inhibition and patterning of Senders and Receivers in the Notch-Delta system. (A)** *Schematic of the model proposed by Collier and collaborators in a system of two cells* [134]. *Pointing arrows indicate Delta-mediated activation of Notch in neighboring cells, while t-shaped arrows indicate Notch-mediated inhibition of Delta within the same cell. (B) On a hexagonal lattice, Notch-Delta lateral inhibition gives rise to a pattern where high-Delta Senders (green hexagons) are surrounded by high-Notch Receivers (red hexagons).*

3.3.3 Variable cell shape as an imprint to guide lateral inhibition

Notch receptors and ligands become available for signaling only once they reach the cell membrane. Therefore, the signaling between pairs of neighbors depends on the geometry of cell-cell contact and their shared membrane area. From a theoretical perspective, this observation offers the possibility to integrate aspects of cell biophysics into the 'standard' Notch-Delta signaling.

Shaya and collaborators [137] investigated the relation between cell size and cell fate by integrating experimental and computational methods. By introducing reports to track the activity of Notch and Delta, they showed that signaling between pairs of nearest neighbors correlates with their shared contact area. Including the role of cell shape requires a more detailed model that explicitly describes ligand-receptor binding as well as transcriptional regulation by the NICD thereafter [132,137–139]. This can be achieved by directly generalizing the model of eqs. (59) to explicitly include variable cell shapes. In this generalized model, a cell is described by a set of three ordinary differential equations for Notch, Delta and NICD

$$\frac{dN_p}{dt} = g_N \frac{1 + \lambda_N \left(\frac{I_p}{I_0}\right)^{n_N}}{1 + \left(\frac{I_p}{I_0}\right)^{n_N}} - kN_p \sum_{j \in N(p)} I_{pj} D_j - \mu N_p, \quad (60a)$$

$$\frac{dD_p}{dt} = g_D \frac{1 + \lambda_D \left(\frac{I_p}{I_0}\right)^{n_D}}{1 + \left(\frac{I_i}{I_0}\right)^{n_D}} - kD_p \sum_{j \in N(p)} I_{pj} N_j - \mu N_p, \quad (60b)$$

$$\frac{dI_p}{dt} = kN_p \sum_{j \in N(p)} D_j - \mu_I I_p. \quad (60c)$$

In eqs. (60a-b), the production of Notch and Delta is regulated by Hill functions that depend on the level of NICD, with $\lambda_N > 1$ and $0 < \lambda_D < 1$ to capture transcriptional activation of Notch and transcriptional inhibition of Delta. Moreover, this model considers binding of Notch and Delta molecules of a given cell ($p$) to ligands and receptors of neighboring cells ($j$); here, $k$ represents a bimolecular binding rate constant between a ligand and a receptor at the surfaces of neighboring cells, and the summation spans over the nearest neighbors of cell ($p$). Binding of Notch with external ligands leads to release of NICD and degradation of the remaining molecular complex; therefore, the binding term results in a loss term in eq. (60a) and a corresponding production term in NICD's equation (60c). Furthermore, the binding term of Notch receptors with external ligands in eq. (60a) is weighted by a contact area term ($I_{pj}$) that describes the shared cell contact area between cells $p$ and $j$. The limit of a regular lattice is achieved by imposing that all weightage terms are equal ($I_{pj} = 1/c$).

On a disordered lattice with variable cell sizes, smaller cells with smaller contact area tend to acquire a Sender state while larger cells tend to acquire a Receiver state [137]. In the perfectly regular lattice, Senders are selected from a homogeneous initial condition simply due to spontaneous breaking of symmetry and amplification of small initial differences [134]. Instead, variation of cell size bias cell fate selection by weighting the amount of signaling between pairs of neighboring cells, and can be thus viewed as an imprint that guides the patterning by breaking the symmetry between cells. This prediction was directly validated in the context of chicken inner ear development, where smaller cells produce Delta at a high rate and eventually become hair cells, while larger cells generally committed to a non-hair, supporting phenotype [137,140].

### 3.3.4 Notch-Jagged signaling guides a transition between lateral inhibition and lateral induction

So far, we have reviewed models that primarily focus on Notch-Delta lateral inhibition. Ligands of the Jagged type, however, can give rise to a positive feedback between neighbors and thus promoting lateral induction of the (high Notch, high Jagged) hybrid Sender/Receiver phenotype. The conflicting effects of Delta and Jagged ligands on cell fate raise interesting questions about their competition in multicellular models where many cells collectively converge to diverse patterns. Boareto and collaborators [139,141] proposed a model of Notch-Delta-Jagged signaling based on the circuit schematic of Figure 13A that directly generalizes the Notch-Delta circuit of eqs. (60a-c) to include NICD transcriptional activation of Jagged and Notch-Jagged binding:

$$\frac{dN_i}{dt} = g_N \frac{1 + \lambda_N \left(\frac{I_i}{I_0}\right)^{n_N}}{1 + \left(\frac{I_i}{I_0}\right)^{n_N}} - kN_i \sum_{j \in N(i)} (D_j + J_j) - \mu N_i, \quad (61a)$$

$$\frac{dD_i}{dt} = g_D \frac{1 + \lambda_D \left(\frac{I_i}{I_0}\right)^{n_D}}{1 + \left(\frac{I_i}{I_0}\right)^{n_D}} - kD_i \sum_{j \in N(i)} N_j - \mu N_i, \quad (61b)$$

$$\frac{dJ_i}{dt} = g_J \frac{1 + \lambda_J \left(\frac{I_i}{I_0}\right)^{n_J}}{1 + \left(\frac{I_i}{I_0}\right)^{n_J}} - kJ_i \sum_{j \in N(i)} N_j - \mu J_i, \quad (61c)$$

$$\frac{dI_i}{dt} = kN_i \sum_{j \in N(j)} (D_j + J_j) - \mu_I I_i. \quad (61d)$$

In this model, a cell behaves as a three-way switch that can assume a Sender, Receiver, or hybrid Sender/Receiver state based on initial conditions, parameters and state of the neighbors. In the limit of a dominant Notch-Delta signaling ($g_D \gg g_J$), this model recovers a lateral inhibition pattern with alternated Senders and Receivers. In the opposite limit of a dominant Notch-Jagged signaling ($g_D \ll g_J$), however, there is a homogeneous solution where all cells are hybrid Sender/Receivers with (high Notch, high Jagged).

Therefore, a single cell exposed to a fixed level of external Notch receptors and ligands can be monostable S, R or S/R, or fall in a regime of multistability based on the levels of external Delta and Jagged (Figure 13B). In a two cells scenario, this model undergoes a sharp transition from lateral inhibition to lateral induction triggered by an increasing production rate of Jagged (Figure 13C). This transition has been used to explain how TNF-$\alpha$, and inflammatory signal that activates Jagged, prevents physiological angiogenesis that relies on Notch-Delta lateral inhibition [142]. On a one-dimensional chain of cells with periodic boundary conditions, intermediate conditions where both Notch-Delta and Notch-Jagged 'modes' of the signaling are relevant ($g_D \approx g_J$) give rise to disordered configurations with mixtures of Senders, Receivers and hybrid Sender/Receivers [139] reminiscent of partially disordered configurations in a spin system (Figure 13D).

Interestingly, both mathematical models and experimental observations suggest a dual role for Jagged. While a strong Notch-Jagged signaling promotes homogeneous patterns of hybrid Sender/Receiver cells, as observed during inner ear development and angiogenesis, a weaker expression of Jagged assists Notch-Delta signaling to organize a precise lateral inhibition. In the context of inner ear development, Petrovic and collaborators [143] showed experimentally that Jagged ligands help refine the pattern of Senders and Receivers by competing with Delta over Notch receptors. This leads to even more NICD, and thus an even stronger inhibition of Delta, in the Receivers. Similarly, a weak activation of Jagged improves

Notch-Delta-driven angiogenesis in an *in vitro* model developed by Kang and collaborators [142].

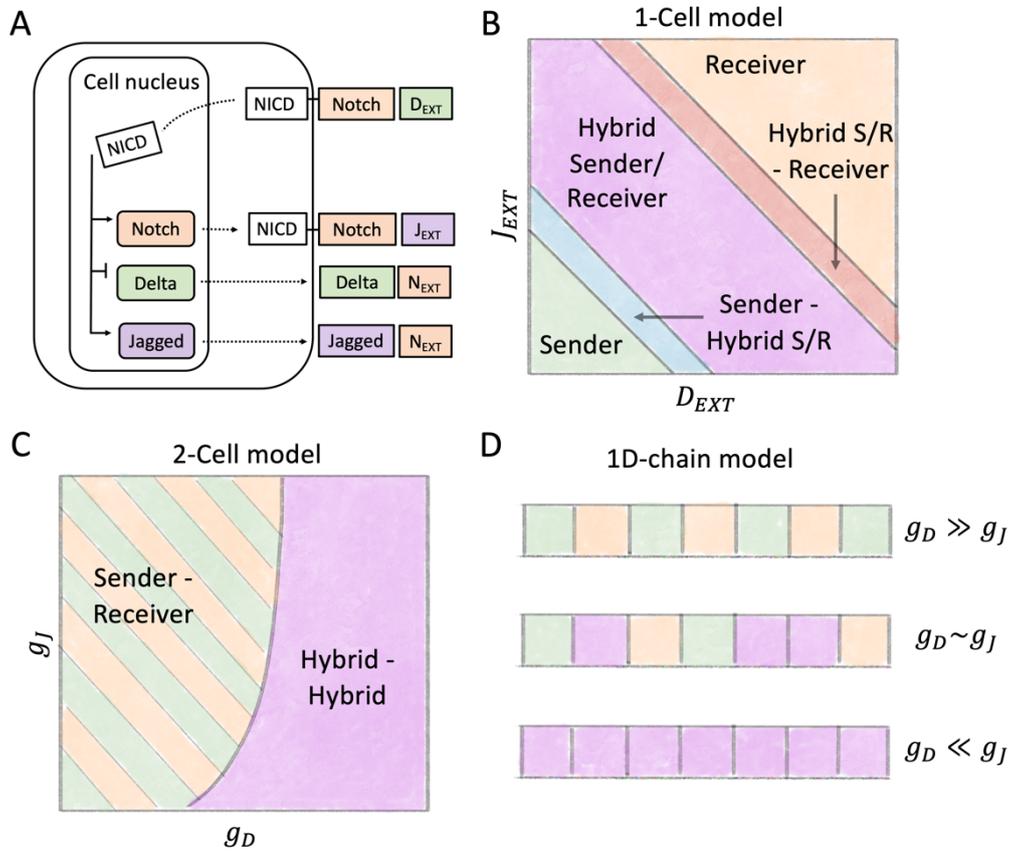

*Figure 13.* **Lateral inhibition and lateral induction patterns in the Notch-Delta-Jagged model. (A)** *Schematic of the Notch-Delta-Jagged circuit in the model of Boareto and collaborators [139]. Dotted arrows summarize production and transport to cell membrane of Notch, Delta and Jagged molecules.* **(B)** *Phase diagram of single cell exposed to constant external levels of Delta and jagged ligands ($D_{EXT}$ and $J_{EXT}$).* **(C)** *Phase diagram of a 2-cell system as a function of cellular production rate of Jagged and Delta ($g_J$ and $g_D$).* **(D)** *Typical patterns observed in a 1D-chain of cells with periodic boundary conditions. Green, orange and purple squares indicate Sender, Receiver and hybrid Sender/Receiver cells, respectively. Panels B, D are adapted from Boareto and collaborators [139]; panel C is adapted from Kang and collaborators [142].*

3.3.5 Stochastic fluctuations lead to optimal lateral inhibition patterning

While the standard paradigm of Notch-Delta lateral inhibition assumes a precise patterning of alternated Senders and Receivers, it is reasonable to assume that spatial constraints might lead to frustrated patterns, in analogy with the relaxation of spin systems. Evidence of patterning mistakes was recently quantified in the context of Tip-Stalk differentiation during sprouting angiogenesis, where Tip cells are occasionally separated by multiple Stalk cells [144,145]. Galbraith and collaborators studied the relaxation of a multicell Notch-Delta system under the effect of white and shot noise, respectively [135]. First, the authors showed that a deterministic Notch-Delta multicell model on a square lattice equilibrates to the "standard", lateral inhibition alternate pattern only for a small set of well-defined initial conditions, such as the "seeding" of a single Sender cell, while typically reaching disordered patterns when starting from more randomized initial conditions. By quantifying the patterning order based on the fraction of correct Sender-Receiver contacts, they demonstrated that intermediate levels of stochastic fluctuations help achieving more ordered patterns, independently of the specific type of noise. The authors suggested an analogy with the navigation of rugged energy landscapes of spin glass systems. The highly ordered salt-and-pepper configuration can be interpreted as a low energy attractor, or global minimum, in a complex, high-dimensional landscape. Conversely, more disordered configurations can be interpreted as local minima with higher energy. Therefore, the relaxation of the Notch-Delta systems is characterized by two timescales. First, on a shorter timescale, the equilibration of the chemical reaction terms leads to the closest local minimum. Second, on a longer timescale, stochastic fluctuations allow a more thorough exploration of the landscape, finally leading to the global minimum with ordered pattern. Intermediate stochastic fluctuations are key to achieve an ordered pattern because low noise levels are not sufficient to escape local minima and navigate the landscape and, conversely, strong fluctuations in the high noise regime become larger than the typical barrier height separating attractors, thus preventing relaxation toward any specific attractor [135].

## 4. Final remarks and future challenges

In this review, we have provided a general overview of modeling and computational strategies to study the dynamics of gene regulatory networks at various scales and levels of detail, ranging from stochastic simulations of individual chemical reactions up to coarse-grained Boolean descriptions of large networks. In particular, our main goal was to showcase tools to study complex biochemical systems that can give rise to multistability, or, in biological terms, the coexistence of multiple states that can be associated with different cell phenotypes. We urge to point out that there is not an intrinsically better approach, but rather different biological questions require choosing the most suitable tools. For example, stochastic models based on the chemical master equation might be suitable to investigate the dynamics of smaller, well-defined circuits where prior knowledge is available about the reaction parameters. More coarse-grained approaches such as continuous models, parameter randomization or Boolean circuits might be more appropriate to study the emerging dynamics of larger circuits where an informed guess of the model's parameters is unfeasible. Throughout the review, we have demonstrated how some of these strategies can be implemented using small, archetypical systems, such as a single transcribing gene and the bistable toggle switch.

Furthermore, we have provided three specific examples to showcase how the "basic" biological building blocks including transcription and translation can be complemented and generalized to include additional biological processes. In the case of the epithelial-mesenchymal transition, mathematical models that only focus on the transcriptional interactions between genes and transcription factors might not be sufficient to fully capture the biology, as post-translational regulation of non-coding RNAs follow a different dynamic and thus requires different mathematical formulations (see section 3.1 and insight box 3). Moreover, the discussion of stem cell differentiation provided an example of how methodologies originally developed for more traditional physical problems, such as quantum mechanics, can be successfully applied to biology, in this case to capture the stochastic fluctuations of an auto-regulatory gene (see section 3.2 and insight box 4). Finally, while most of the existing modeling efforts tend to focus on the dynamics of individual cells, the discussion of Notch signaling showed how these models can be applied to multicell, spatial models, thus raising interesting connections between regulation of cell fate and spatial patterning (see section 3.3).

In conclusion, we stress that, while existing methodologies reviewed here and elsewhere provide an exhaustive framework to describe gene regulation, many open questions and challenges lie ahead. For example, the fast development of transcriptomics methods has radically improved our resolution on gene regulation at the single cell level [25]. Single cell RNA sequencing (scRNA-seq) allows to estimate the number of transcripts of each RNA species in individual cells[81]. Some computational methods have been proposed recently to infer the interactions between genes from scRNA-seq data by coupling modeling and statistical regression [28,82,146]. Furthermore, it is becoming increasingly clear that the decision-making of cell fate specification does not solely rely on regulatory interactions between genes. For example, in both the cases of EMT and Notch signaling, phenotypic transitions also imply changes in the cell's mechanical properties, which in turn regulate the transcriptional signaling, thus giving rise to mechano-chemical feedbacks[96]. Therefore, the integration of chemical and mechanical regulation in cell-fate specification represents a novel and intriguing challenge for theoretical and computational modeling. Tackling these exciting open questions will require an even stronger combination of existing and new physical and mathematical concepts towards the description of complex biological systems.

**Funding statement**

This project was supported by grants from National Institutes of Health grants U01AR073159 (Q.N.) and R01AR079150 (Q.N.), National Science Foundation grants DMS1763272 (Q.N.) and MCB2028424 (Q.N.), and a Simons Foundation grant (594598 to Q.N.). Work at the Center for Theoretical Biological Physics sponsored by the NSF (Grant PHY-2019745 and PHY-2210291). JNO is CPRIT Scholar in Cancer Research sponsored by the Cancer Prevention and Research Institute of Texas. MKJ was supported by Ramanujan Fellowship (SB/S2/RJN-049/2018) awarded by Science and Engineering Research Board, Department of Science and Technology, Government of India.

**Author contribution**

F.B. and D.J. wrote the manuscript. All authors edited the manuscript. Q.N, M.K.J., and J.N.O. supervised the research.